\documentclass[sigconf]{acmart}

\usepackage{algorithm}
\usepackage{algorithmic,eqparbox}
\usepackage{multirow}
\usepackage{color}
\usepackage{balance}
\usepackage{url}
\usepackage{xspace}
\usepackage{mathtools}
\usepackage{enumitem}
\usepackage{subcaption}
\usepackage{placeins}
\usepackage[toc,page]{appendix}
\newcommand{\ie}{\emph{i.e.,}\xspace}
\newcommand{\eg}{\emph{e.g.,}\xspace}
\newcommand{\etal}{\emph{et al.}\xspace}
\newcommand{\eat}[1]{}

\def\EndOfProof{\nolinebreak\ \hfill\rule{1.5mm}{2.7mm}}

\AtBeginDocument{%
  }

\setcopyright{none}

\begin{document}

\title{PANACEA: Towards Influence-driven Profiling of Drug Target Combinations in Cancer Signaling Networks}

\author{Baihui Xu}
\affiliation{%
  \institution{Nanyang Technological University}
  \country{Singapore}
}
\email{nerissa.xu@ntu.edu.sg}

\author{Sourav S Bhowmick}
\affiliation{%
  \institution{Nanyang Technological University}
  \country{Singapore}
}
\email{assourav@ntu.edu.sg}

\author{Jiancheng Hu}
\affiliation{%
  \institution{National Cancer Centre Singapore}
  \country{Singapore}}
\email{hu.jiancheng@nccs.com.sg}

\renewcommand{\shortauthors}{Xu, Bhowmick, and Hu.}

\begin{abstract}

Data profiling has garnered increasing attention within the data science community, primarily focusing on structured data. In this paper, we introduce a novel framework called \textsc{panacea}, designed to \textit{profile} known cancer target combinations in cancer type-specific signaling networks. Given a large signaling network for a cancer type, known targets from approved anticancer drugs, a set of cancer mutated genes, and a \textit{combination size parameter} $k$, \textsc{panacea} automatically generates a \textit{delta histogram} that depicts the distribution of $k$-sized target combinations based on their \textit{topological influence} on cancer mutated genes and other nodes. To this end, we formally define the novel problem of \textit{influence-driven target combination profiling} ($i$-TCP) and propose an algorithm that employs two innovative personalized PageRank-based measures, \textit{PEN distance} and \textit{PEN-diff}, to quantify this influence and generate the delta histogram. Our experimental studies on signaling networks related to four cancer types demonstrate that our proposed measures outperform several popular network properties in profiling known target combinations. Notably, we demonstrate that \textsc{panacea} can significantly reduce the candidate $k$-node combination exploration space, addressing a longstanding challenge for tasks such as \textit{in silico} target combination prediction in large cancer-specific signaling networks.

\end{abstract}

\maketitle

\section{Introduction}\label{sec:intro} 
\begin{quote}
\textit{``Improving the quality of target selection is widely considered as the single most important factor to improve the productivity of the pharmaceutical industry.''}

\hspace{45mm}\textit{Csermely \etal}~\cite{CP13}
\end{quote}

Cell signaling is a complex network characterized by interactions between various signaling pathways, making it essential to understand signal flow, as disruptions can lead to numerous diseases~\cite{Wein07}. For instance, mutations in genes encoding key signaling proteins like \texttt{RAS} and \texttt{PI3K} are frequently associated with various cancers.

The approach to treating these altered signaling pathways has long been dominated by the ``one-target, one-drug'' paradigm, which focuses on finding a single chemical entity that interacts with a specific target. A \textit{target} is an endogenous factor, typically an enzyme or receptor, that affects the outcome of a disease or medical condition\footnote{\scriptsize  In pathogen-related diseases, the target can sometimes be endogenous to the pathogen, instead of the host. In this research, our focus is on non-pathogen-related diseases.}. However, most complex diseases are \textit{polygenic}, involving a network of interacting genes and their products rather than a single gene. This realization has shifted attention towards \textit{combination therapy}, which aims to simultaneously target multiple molecules within a disease-related signaling network~\cite{CP13,HW11,Mitchell03,Wein07}.

Combination therapy has the potential to provide greater benefits than monotherapy by achieving comparable efficacy at lower doses through synergistic effects, while also reducing the likelihood of resistance developing~\cite{Mitchell03}. Consequently, identifying effective target combinations for specific diseases (a.k.a \textit{multi-target selection}~\cite{HW11}) has become a recognized challenge in the field~\cite{CP13,HW11,Mitchell03}. This has led to increasing research focused on developing \textit{in silico} techniques for predicting target combinations~\cite{chua2017synergistic,HC+19,Alvarez16,TP21,WS13,li18,Chua2012,timma}. However, existing studies face two significant limitations (see Appendix~\ref{app:tcd} for details). First, they do not take into account the characteristics (\ie \textit{profiles}) of \emph{approved} target combinations for specific diseases  to guide discovery. Second, they are primarily designed for small networks, as it is computationally intractable to exhaustively explore the entire candidate combination space in large networks. This paper presents a novel framework designed to \textit{profile} known target combinations in cancer signaling networks, which has the potential to address these limitations.

\vspace{-1ex}
\subsection{Profiling Known Target Combinations} 
\textit{Data profiling} is the set of activities and processes to determine the metadata of a given dataset. In~\cite{wiki}, it is defined as \textit{``Data profiling is the process of examining the data available in an existing data source [...] and collecting statistics and information about that data.''} It has been widely studied in the context of structured data~\cite{AG+18}. The type of data profiling we undertake is typically informed by the downstream analysis we aim to conduct.

\begin{figure*}[t]
    \centering
         \includegraphics[width = \linewidth]{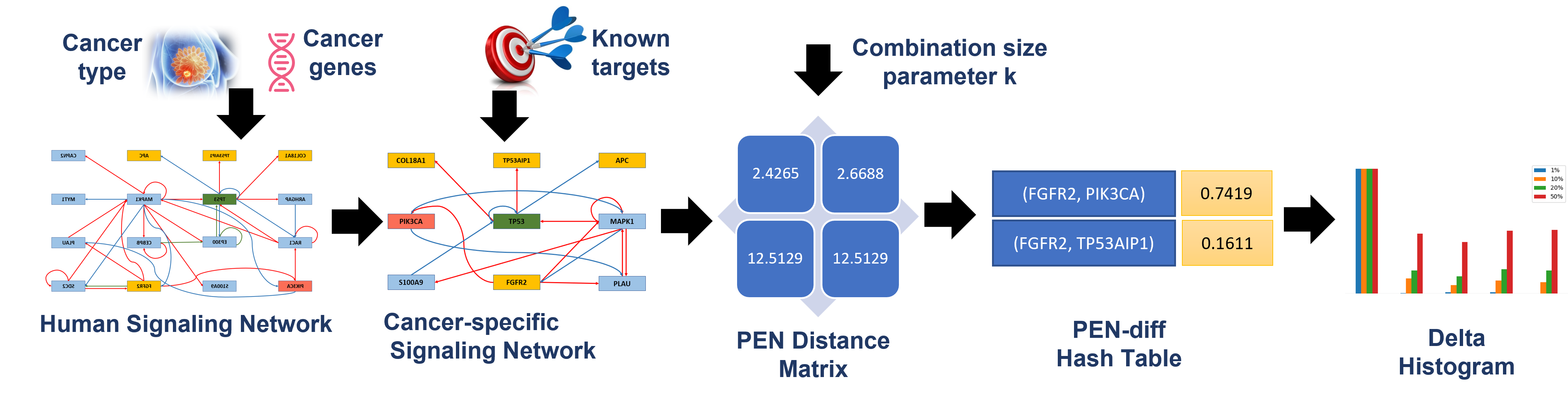}
         \vspace{-2ex} \caption{[Best viewed in color] The PANACEA framework.} \label{fig:overview}
\vspace{-1ex}\end{figure*}

The \textit{Cancer Drugs Database}, maintained by the \textit{Anticancer Fund}~\cite{acf-drugs-db}, offers a curated list of anticancer drugs approved by one or more regulatory agencies (such as the FDA or NCI) for various cancer types, along with their associated target combinations. In this paper, we aim to profile these target combinations in the signaling network of a specific cancer type. Instead of focusing on individual targets, we profile known target combinations from approved drugs, presenting a more accurate view of the topological features of cancer target combinations in large signaling networks. These profiles potentially create opportunities for judiciously \textit{selecting} candidate combinations in a \textit{profile-driven} manner. They can serve as a guide for determining which candidate $k$-node combinations to further analyze for target combination prediction (\ie $k$ molecules that can be targeted simultaneously). For example, one might choose to investigate network regions where most or very few known target combinations are located.\eat{ Additionally, profiling may enhance the discovery of novel drug combinations by mapping known target combinations to their corresponding drugs.}

\vspace{-1ex}
\subsection{Overview \& Contributions} 
Profiling known target combinations in signaling networks presents a significant challenge. Since various profiles can be generated, not all are effective for characterizing these combinations to address downstream problems such as target combination prediction. For example, degree or PageRank distributions may fail to capture the off-target effects of these combinations, which are a primary reason for drug failures~\cite{Vogel07}. Therefore, it is crucial to select appropriate features for profiling that can support the downstream analyses. Additionally, to manage large and noisy cancer-specific signaling networks, the chosen features must be purely topological. This means they should not rely on complete mathematical models of the underlying cancer network, nor on the pharmacological or chemical properties of all involved molecules, or the genomic and proteomic data of patients. Instead, the profiling strategy must adopt a realistic perspective, acknowledging that such data is often unavailable in many areas of cancer signaling networks.

In this paper, we introduce a novel \textit{influence-driven target combination profiling} ($i$-TCP) problem to tackle these challenges. Given a large signaling network $G_C$ for a cancer type $C$ (\eg breast, colorectal), known targets tackled by anticancer drugs of $C$, a set of cancer mutated genes (\eg oncogenes) in $G_C$, and a \textit{combination size parameter} $k> 1$,  the goal of $i$-TCP is to generate a \textit{delta histogram} that depicts the distribution of $k$-size known target combinations in $G_C$ based on their \textit{topological influence} (\ie strength of connection) on cancer mutated genes and rest of the nodes in $G_C$. 

We present a novel framework called \textsc{panacea} (\textbf{P}ersonalized p\textbf{A}gerank-based profili\textbf{N}g of c\textbf{A}n\textbf{CE}r t\textbf{A}rgets) to address the $i$-TCP problem (Figure~\ref{fig:overview}). This framework utilizes a new personalized PageRank-based (PPR)~\cite{YW+24} measure called \textit{PEN distance} to assess the topological influence of one node (\eg a drug target) on another (\eg an oncogene) in a cancer signaling network. Intuitively, a smaller PEN distance suggests a greater influence due to a higher number of paths between the node pair. Building on PEN distance, we also propose a measure called \textit{PEN-diff}, which captures the difference in average influence of a $k$-node combination 
on a set of cancer mutated genes compared to the remaining nodes in the network. A positive PEN-diff value indicates that the average influence on the cancer mutated genes is greater than on other nodes, effectively capturing the off-target effects of a drug from a topological perspective (\ie impact on rest of the nodes in $G_C$). Using the PEN-diff values of $k$-node combinations and known target combinations in $C$, we build a \textit{delta histogram} to depict the distribution of known $k$-target combinations in the PEN-diff space of $G_C$. Our experimental results highlight the advantages of PEN-diff-based profiling (\ie influence-driven profiling) of known target combinations compared to several other popular topological measures (\eg PageRank, network distance).

In summary, this paper makes the following contributions. 

\begin{itemize}

\item We present a novel problem termed \textit{influence-driven target combination profiling} to analyze known target combinations in cancer signaling networks, introducing the framework called \textsc{panacea} to tackle this issue. To the best of our knowledge, this is the first effort to specifically profile known target combinations in a disease-related signaling network, bridging the topics of data profiling and combination therapy. 

\item We introduce a new PPR-based measure, PEN-diff, designed to capture the differences in influence of $k$-target combinations on cancer genes compared to other nodes in a cancer signaling network. While PPR computation methods have been extensively studied for their efficiency and scalability, they have not been applied to profiling target combinations.

\item We experimentally demonstrate that \textsc{panacea} outperforms two baseline strategies in profiling target combinations across four different cancer types.

\end{itemize}

The rest of the paper is organized as follows. In Section~\ref{sec:back}, we present relevant background information. We formally introduce the $i$-TCP problem in Section~\ref{sec:tcp}. We present the \textsc{panacea} framework to address it in Section~\ref{sec:panacea}. The performance of \textsc{panacea} is reported in Section~\ref{sec:exp}.  We discuss related work in Section~\ref{sec:rel}. The last section concludes the paper\eat{ by highlighting intriguing research challenges for the data management community}.

\vspace{0ex}
\section{Background} \label{sec:back}
In this section, we begin by introducing the network representation of human signaling. Then, we introduce the databases of cancer mutated genes and drug targets that we leverage. Finally, we briefly describe the notion of \textit{personalized PageRank}\eat{ that we shall be exploiting later}.  

\vspace{-1ex}
\subsection{Human Signaling Network}
Cell signaling refers to the process of cellular communication by which proteins and other factors interact to perform multifaceted information processing functions, such as growth, survival, and differentiation. These cell signalings usually occur through different signaling pathways. The interconnections between these signaling pathways are crucial to understand the flow of signaling events as gene mutations or epigenetic changes can alter the cellular signaling events leading to various diseases such as cancer. 

\begin{table}[t] \scriptsize
  \vspace{-2ex}\caption{Features of the reduced signaling network.}
  \label{tb:summary}
  \begin{tabular}{ll} 
   \toprule
   Feature & Value \\ 
    \midrule
    No. of nodes & 6,009  \\ 
    No. of edges & 41,358 \\
    Density & 0.0011 \\
    Average clustering coefficient & 0.0748 \\ 
    No. of strongly connected components & 3,335 \\
   Size of the largest strongly connected component & 2,660\\
  \bottomrule
  \end{tabular}
\vspace{-3ex} \end{table}

A biological signaling network describes the interactions between molecular species (or molecules) in cell signaling. Each interaction takes the form of a biochemical reaction. Large signaling networks often are modeled simply as large graphs. For instance, Cui \etal~\cite{cui2007map} modeled the human signaling network as a graph where nodes represent proteins. Directed links are used to represent activation or inhibition whereas undirected links represent physical interactions of proteins that are not characterized as activating or inhibitory. There are two types of directed links, incoming and outgoing. The incoming link represents a signaling from another node whereas an outgoing link represents a signal to another node. These two types of directed links are collectively referred to as \textit{signal links}.  In contrast, the physical links are referred to as \textit{neutral links}. In this paper, we represent the human signaling network using this model.
 
In our work, we utilize the signaling network used in the study reported in~\cite{zaman2013signaling}. Specifically, Zaman \etal~\cite{zaman2013signaling} collected and curated this network based on the previous studies~\cite{awan2007regulatory,cui2007map,li2012human}. It contains $6,305$ nodes and $62,937$ links. There are $33,398$ activation links (\ie positive links), $7,960$ inhibitory links (\ie negative links), and $21,579$ physical links (neutral links). Since we focus on signal links, we reduce this network by removing all neutral links.  Table~\ref{tb:summary} summarizes the features of the \textit{reduced} network that we shall be using in the sequel. Figure~\ref{fig:eg-reduced} shows a fragment of the reduced network.

\vspace{-1ex}
\subsection{Cancer Mutated Genes} \label{sec:gene}
\begin{figure}[t]
    \centering
         \includegraphics[width = \columnwidth]{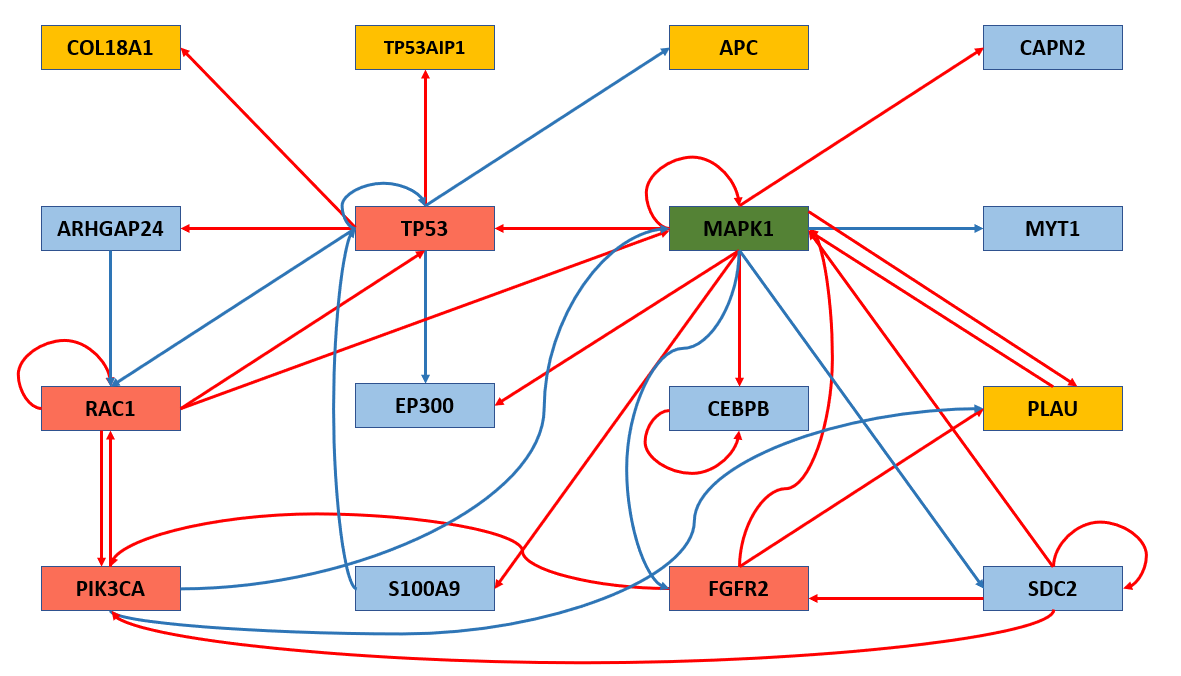}
         \vspace{-3ex} \caption{[Best viewed in color] A small subnetwork of the reduced human signaling network. The red edges are positive links and the blue edges are negative links.  Nodes in yellow are drug targets. Nodes in green are cancer genes. Red nodes are both a cancer gene and a drug target.}
    \label{fig:eg-reduced}
\vspace{0ex}\end{figure}

Genes that, when mutated, result in tumorigenesis often lead to the aberrant activation of certain downstream signaling nodes giving rise to dysregulated growth, survival and/or differentiation. We gather cancer mutated genes from the \textit{\textsf{COSMIC} Cancer Gene Census} database~\cite{sondka2018cosmic}. These genes can be categorized into three types, positive regulators (oncogenes), negative regulators (tumor suppressors), and fusion genes. Since majority of the mutated genes in cancer are oncogenes~\cite{cui2007map},  we focus only on them in this work. In our dataset, there are $318$ oncogenes, $252$ ($79.25\%$) of which can be located in the constructed human signaling network.\eat{ The genes SH3GL1 and CTNNA2 are in the human signaling network but not in the reduced signaling network.} Some example cancer genes are shown in Figure~\ref{fig:eg-reduced}.

\begin{figure*}[t]
    \centering
         \includegraphics[width = 0.24\linewidth]{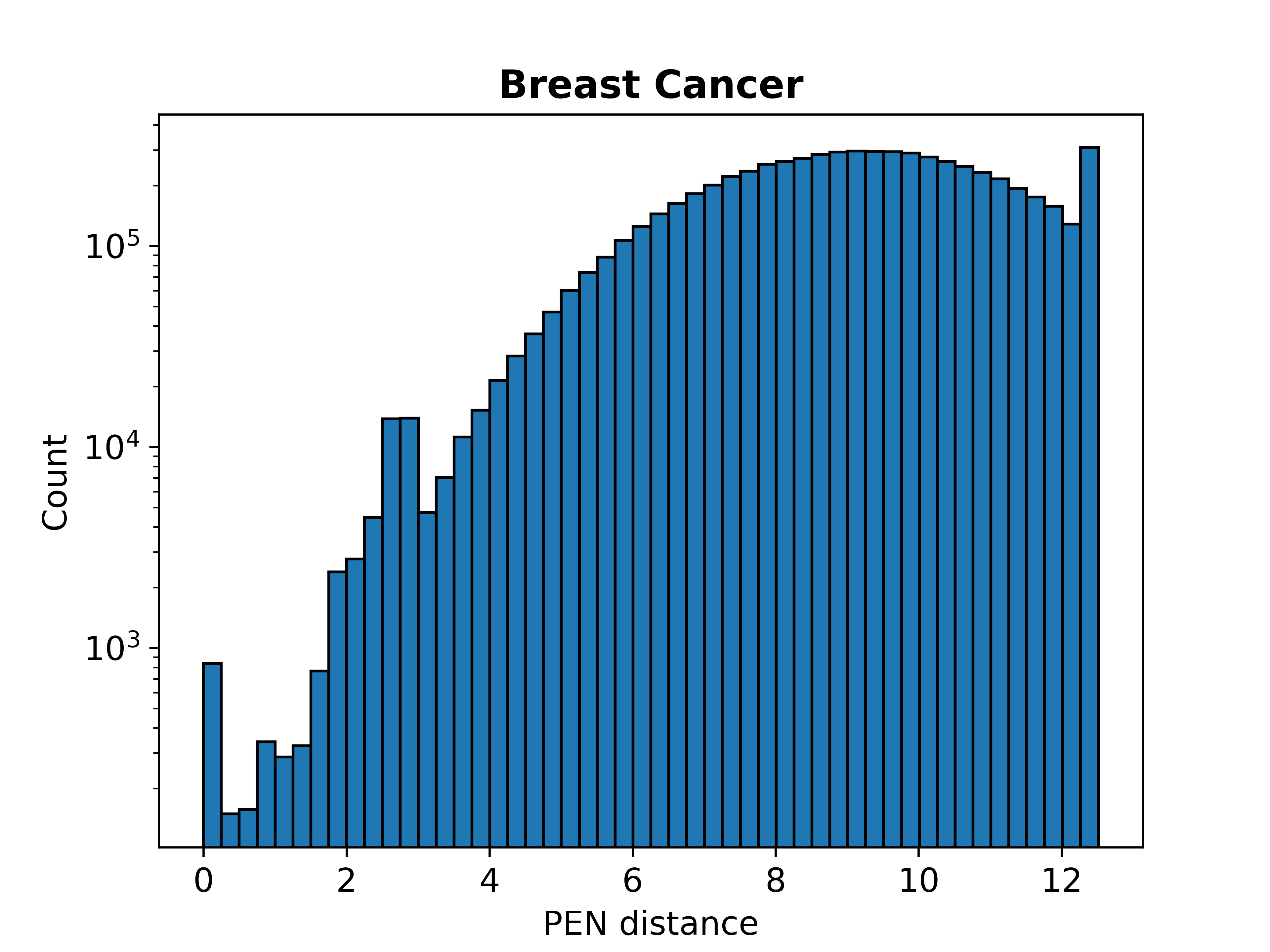}
        \includegraphics[width = 0.24\linewidth]{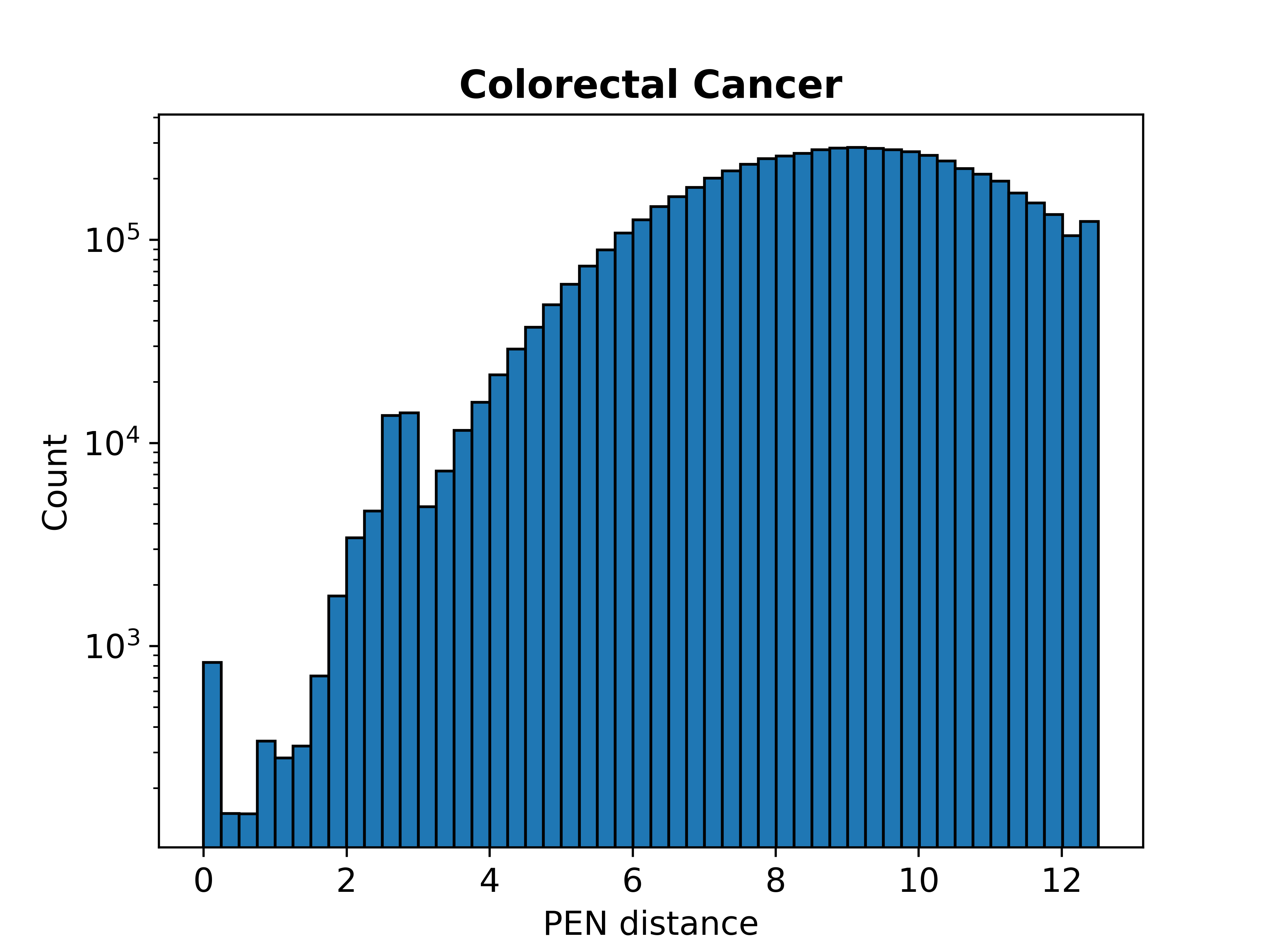}
         \includegraphics[width = 0.24\linewidth]{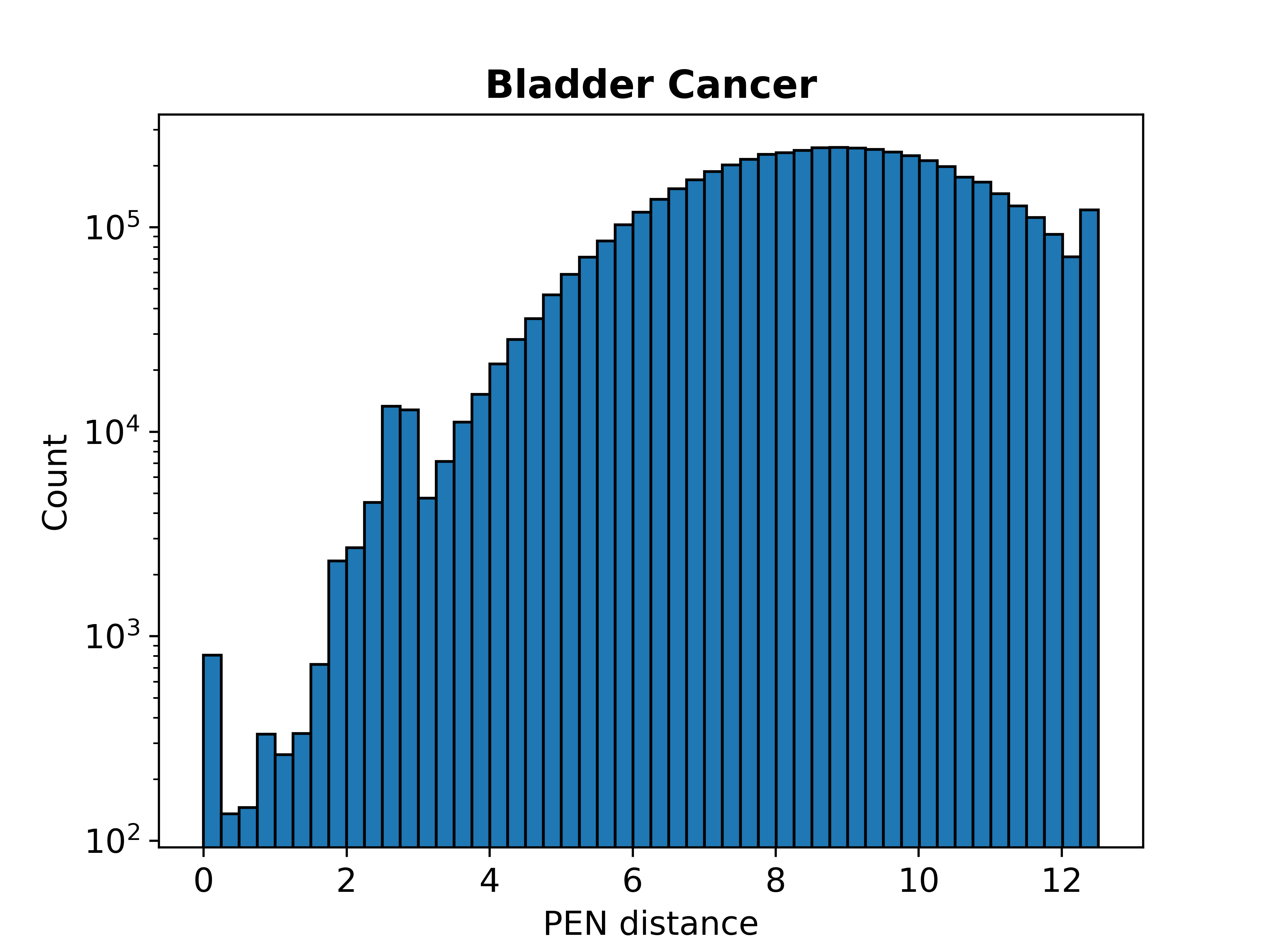}
         \includegraphics[width = 0.24\linewidth]{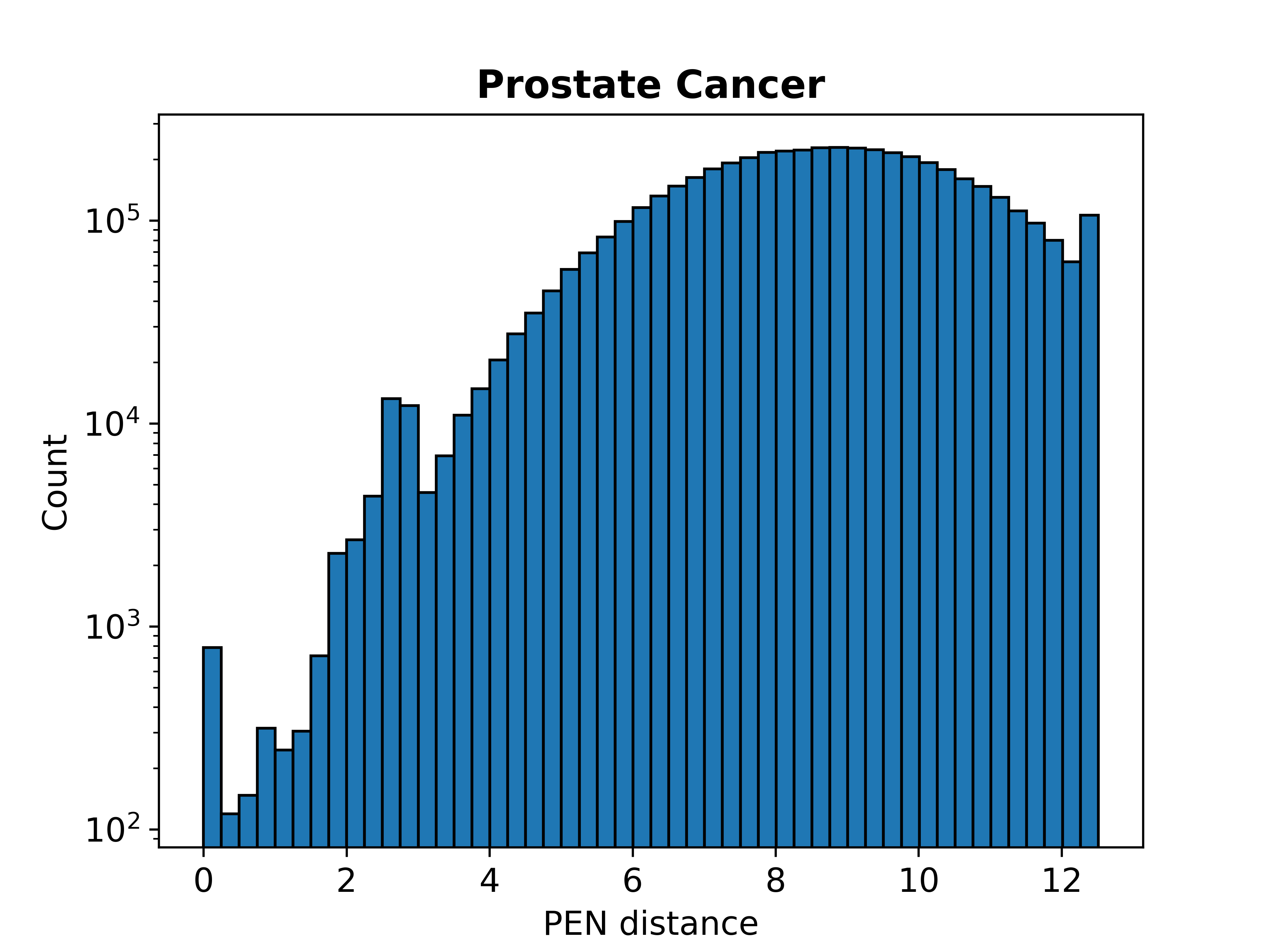}
         \includegraphics[width = 0.24\linewidth]{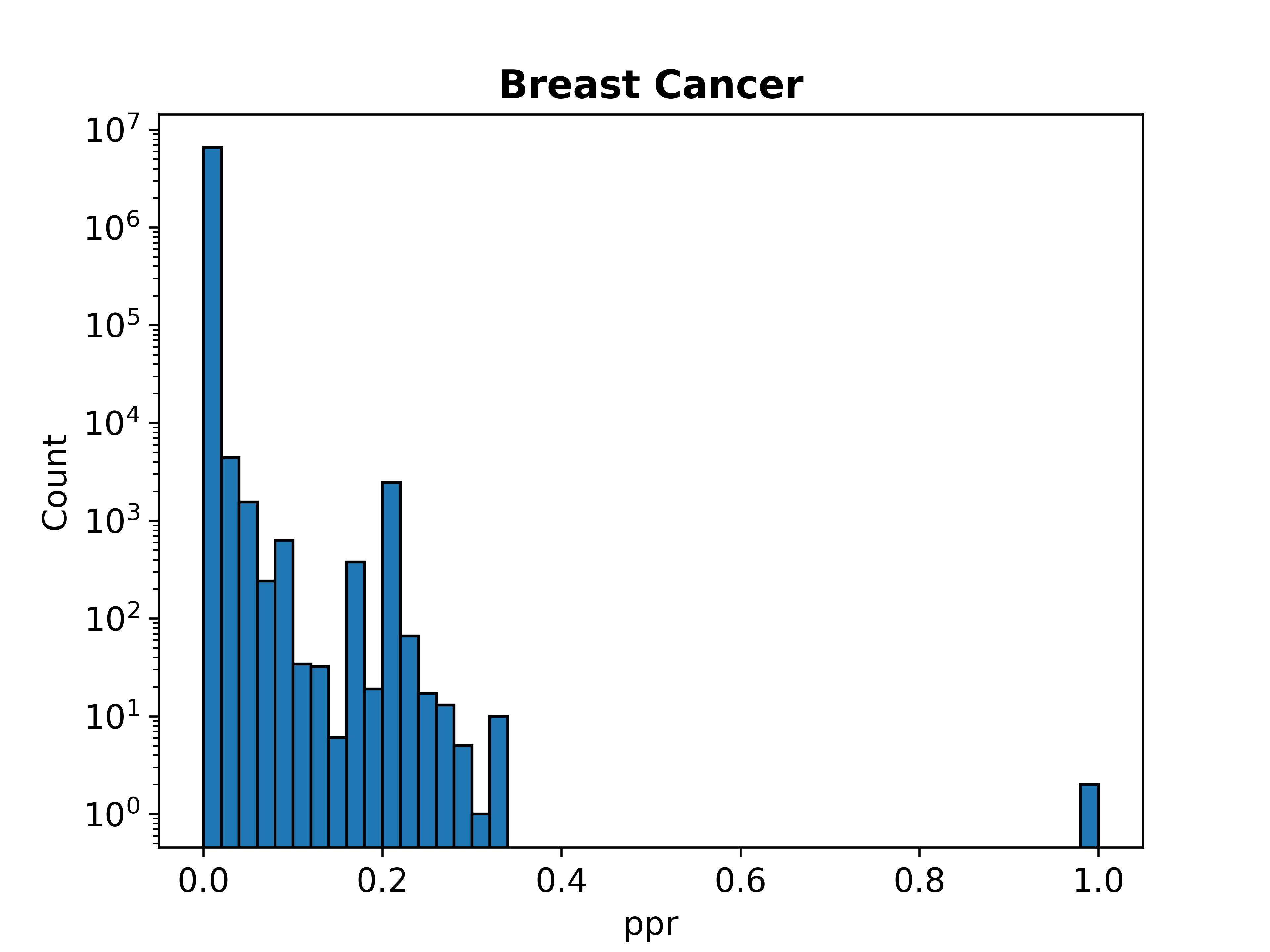}
         \includegraphics[width = 0.24\linewidth]{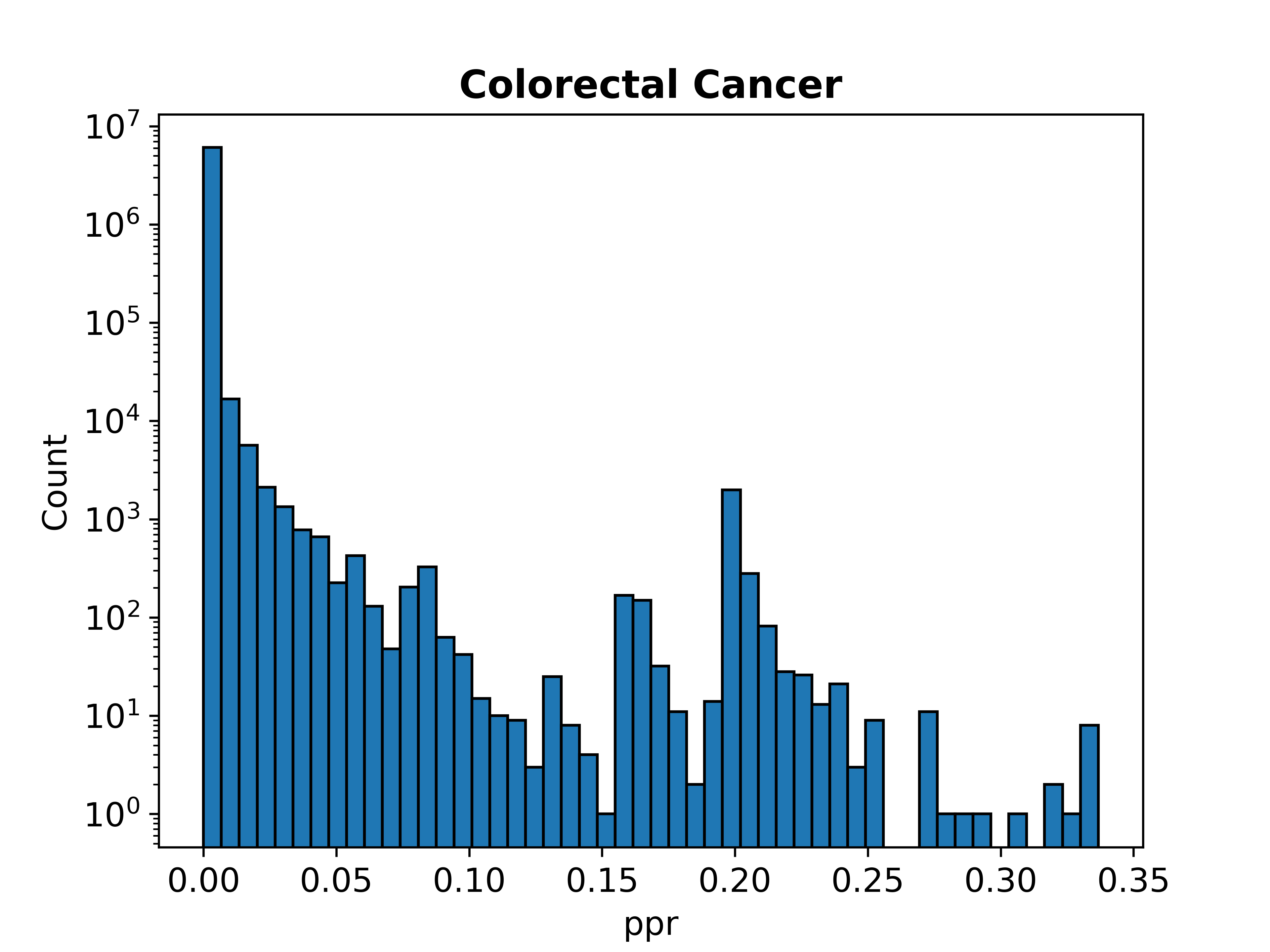}
         \includegraphics[width = 0.24\linewidth]{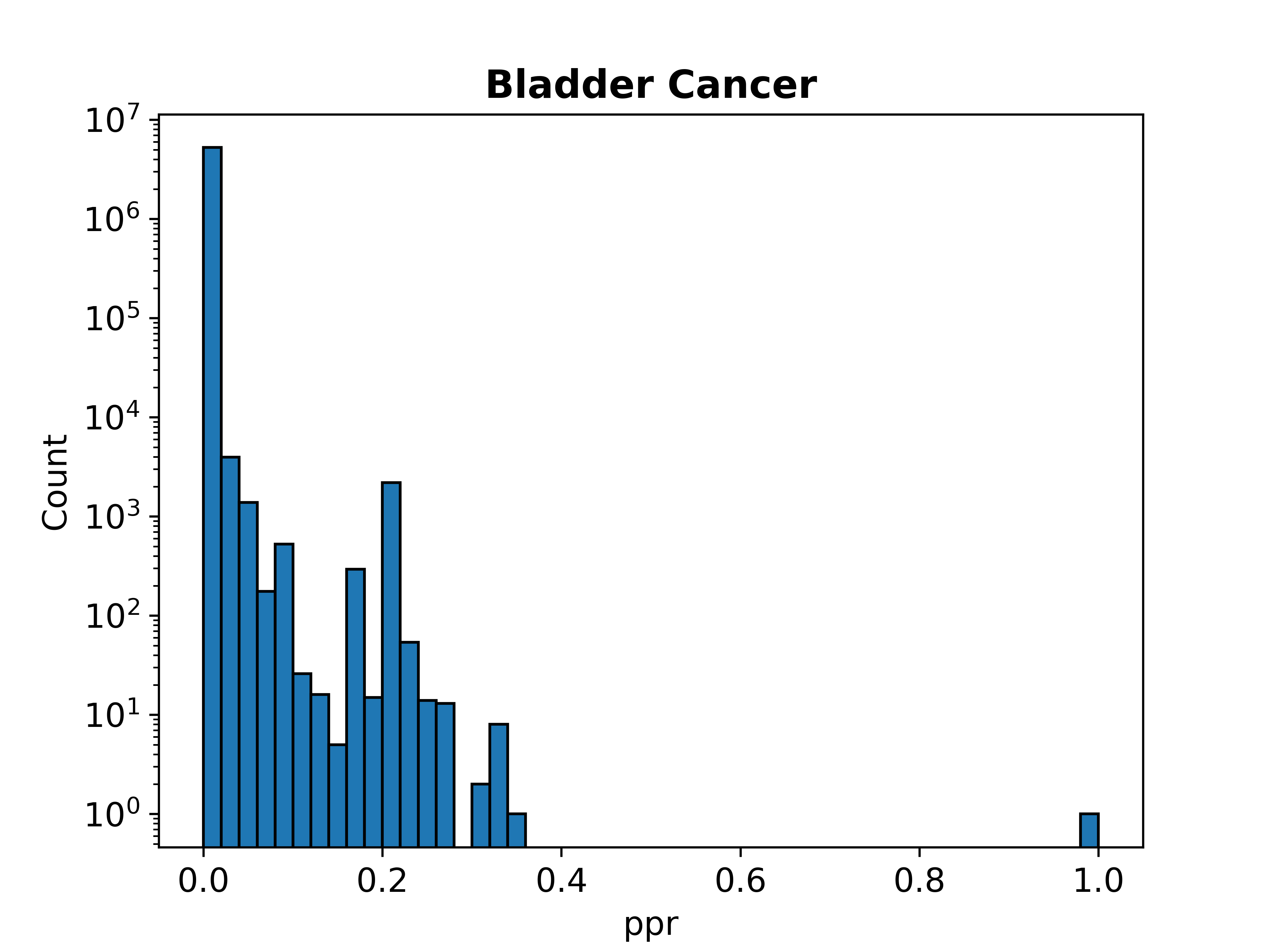}
         \includegraphics[width = 0.24\linewidth]{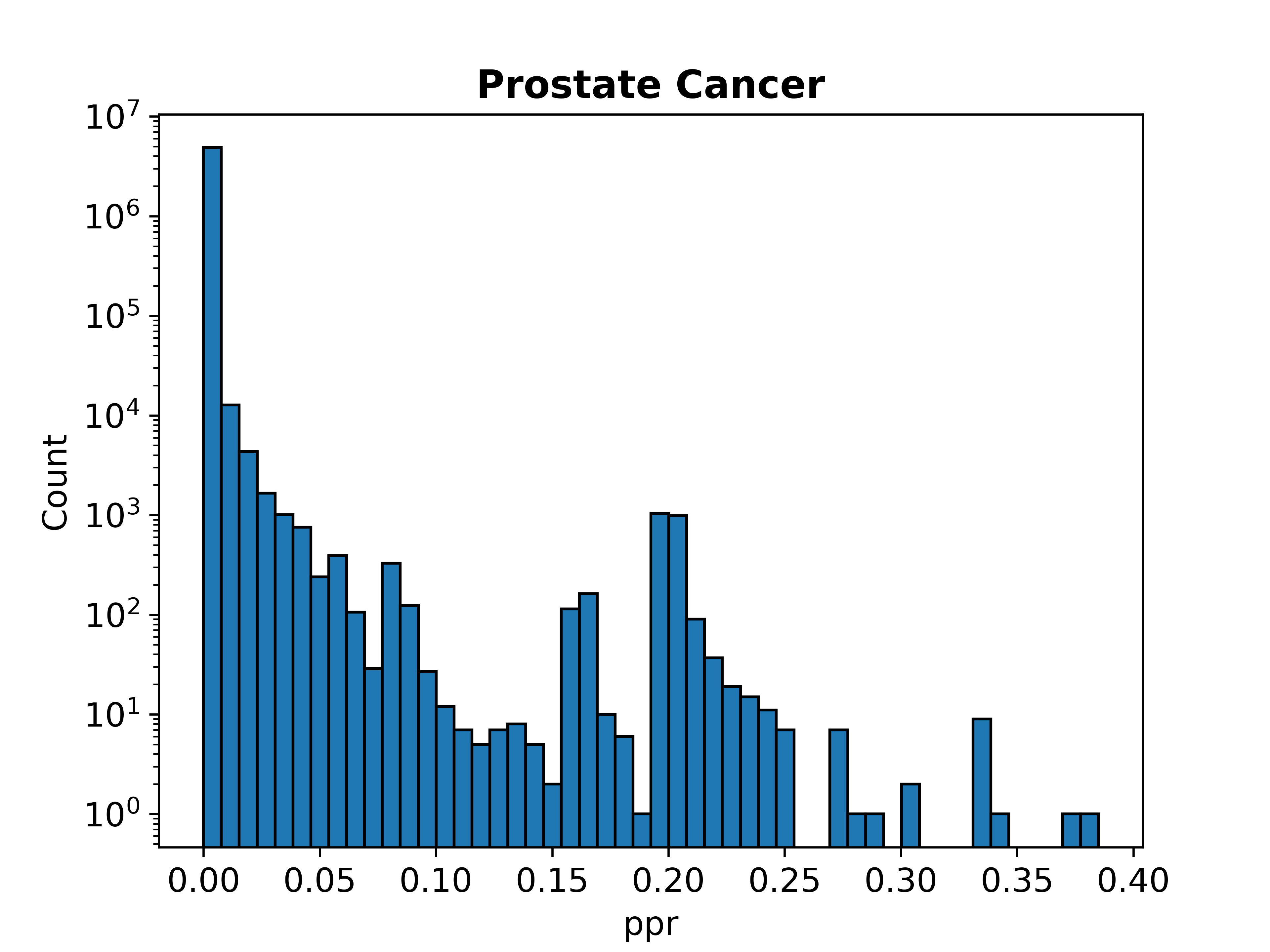}
         \vspace{-2ex} \caption{PEN distance (top row) and PPR (bottom row) distributions of different target-aware cancer-specific signaling networks.}
    \label{fig:pen}
\vspace{0ex}\end{figure*}

\vspace{-1ex}
\subsection{Cancer Drug Targets} \label{sec:target}
A \textit{disease node} in a signaling network is a molecule that is either involved in some dysregulated biological processes implicated in a disease (\eg cancer), or is of interest due to its potential role (\eg oncogene) in the disease. Targeted therapy is the foundation of precision medicine, which is a type of cancer treatment that targets specific nodes (\ie drug targets) that can modulate these disease nodes in order to control the survival, division, and spread of cancer cells.  For example, a disease node in the \textsf{MAPK} signaling network is phosphorylated \textsf{ERK} (\textsf{ERKPP}). Hence, drug targets are those nodes (\eg proteins) that can reduce the activity of \textsf{ERKPP} in order to curb the proliferative potential of cancer cells.

We obtain data on cancer drug targets from the \textit{Cancer Drugs Database} maintained by \textit{Anticancer Fund}~\cite{acf-drugs-db}. This database provides a curated list of anticancer drugs approved by one or more regulatory agencies (\eg FDA or NCI) as well as their targets. Among the list of $1,025$ drug targets, $119$ ($11.61\%$) are oncogenes and $725$ ($70.73\%$) are in the human signaling network \cite{acf-drugs-db}. Some example drug targets are depicted in Figure~\ref{fig:eg-reduced}.

\vspace{-1ex}
\subsection{Personalized PageRank (PPR)}  \label{sec:ppr}
Intuitively, the \textit{personalized PageRank} (PPR)~\cite{YW+24} is a classic metric that measures the proximity of nodes in a graph based on how well they are connected to each other via random walks. That is, PPR of a node $t$ in a directed graph reflects its relevance from the perspective of a source node $s$.  Formally, let $G = (V, E)$ be a directed graph where $V$ is the set of nodes (vertices) and $E$ is the set of edges (links). Given a source node $s\in V$ and a \textit{jump factor} $\alpha$, a walker starts from $s$ to traverse $G$, and at each step, the walker either (a) terminates at the current node with a probability $\alpha$, or (b) jumps to a randomly selected out-neighbor of the current node. For any node $t \in V$, the \textit{personalized PageRank} (PPR) $\pi(s, t)$ is the probability that a random walk (RW) from $s$ terminates at $t$. Intuitively, a large $\pi(s, t)$ indicates that many paths exist from $s$ to $t$. That is, $s$ is well connected to $t$.  The $\alpha$ value is usually set to 0.15 or 0.2~\cite{YW+24,ZY+23}.

\vspace{0ex} 
\section{The $i$-TCP Problem} \label{sec:tcp}  
In this section, we formally define the \textit{influence-driven target combination profiling} ($i$-TCP) problem that we address in this paper. Intuitively, the goal of the $i$-TCP problem is to generate a \textit{delta histogram} that visually represents the distribution of known $k$-target combinations w.r.t. their \textit{topological influence} in a cancer signaling network. Given a large signaling network $G_C=(V_C, E_C)$ for a cancer type $C$ (\eg breast, colorectal), known targets $X_{t} \subseteq V_C$ of $C$, a set of cancer mutated genes $X_g \subseteq V_C$ in $G_C$, and a $k$-combination node set $h_k$ where $k> 1$ and $h_k \subset V_C$,  the \textit{topological influence} of $h_k$ is defined by a function $f(h_k, X_g, G_C)$ that quantifies the \textit{aggregate} influence of $h_k$ on $X_g$ and $(V_C - X_g)$. 

A \textit{delta histogram} $\mathbb{H}_C$ of $G_C$ is an equi-width histogram whose $X$-axis represents $N_{bucket}$ buckets and the $Y$-axis represents the \textit{percentages} of known $k$-target combinations in $G_C$ in each bucket. A $k$-node combination $h_k$ is assigned to a bucket $b_i=[r_{min}, r_{max}]$ if $r_{min} < f(h_k, X_g, G_C)\leq r_{max}$. The \textit{percentage} of known $k$-target combinations in $b_i$ is the number of known target combinations in the top-$m$ percentage of $k$-node combinations for $m > 0$.  

\begin{definition} \label{def:prob} {\em
 Given a signaling network $G_C = (V_C, E_C)$ for a cancer type $C$, a set of known targets $X_{t} \subseteq V_C$, a set of cancer-mutated genes $X_{g} \subseteq V_C$, a combination size parameter $k>1$, and the number of buckets $N_{bucket} > 1$, the goal of \textit{\textbf{influence-driven target combination profiling}} problem is to compute the following: 
\begin{equation}
    \mathbb{H}_{k,{N_{bucket}}}=\mathcal{F}(\mathcal{G}(G_C,k,X_g), X_t, N_{bucket})
\end{equation}
where $\mathcal{G}(G_C,k,X_g)$ is a function that computes $f(h_k, X_g, G_C)$ of the $k$-combination nodes in $G_C$ and $\mathcal{F}(\cdot)$ is a function that computes an $N_{bucket}$-delta histogram of the known $k$-target combinations in $X_t$. \/}
\end{definition}

\eat{Observe that the problem can be modeled as the optimization of a \textit{constraint satisfaction problem} (\textsc{csp}) which is NP-hard~\cite{}.} In \textsc{panacea} the topological influence function $f(\cdot)$ is realized by a novel PPR-based measure called \textit{PEN-Diff} that captures the interplay between the influence of a $k$-target combinations on $X_g$ and rest of the network (\ie off-target effect). 

\vspace{0ex}
\section{The PANACEA Framework}\label{sec:panacea}
In this section, we present the \textsc{panacea} framework to address the $i$-TCP problem in a cancer signaling network. We begin by introducing a PPR-based node distance measure called \textit{PEN distance} that serves as the foundation for this task.  

\vspace{-1ex}
\subsection{PEN Distance}
Recall that the key goal of our work is to characterize the topological influence (\ie strength of connection) between drug target combinations and cancer genes, as well as other nodes within a cancer signaling network. While several studies in drug and target combination discovery utilize various distance-based measures~\cite{CKB19,HC+19}, it is important to recognize that the shortest path is not the only way a target can impact other nodes. Targets may connect to other nodes via multiple alternative paths. Therefore, assessing topological influence should leverage network propagation-based methods. Cowen \etal~\cite{CI+17} have reviewed several of these methods, which include PPR, RW, and diffusion kernels. 
In our work, we focus on capturing connectivity-based relationships through a PPR-based distance measure
known as \textit{PEN distance} (\textbf{P}ersonalized pag\textbf{E}rank-based \textbf{N}ode distance). It is worth noting that RW and diffusion kernels are unsuitable for calculating PEN distance between node pairs.  In particular, the diffusion kernel serves as the continuous-time equivalent of random walks with restart~\cite{CI+17}.

Intuitively, the PPR value  $\pi(s, t)$ indicates the node $t$'s importance to the node $s$. Recall that if  $\pi(s, t)$ is high then $t$ can be reached from $s$ via many paths (\ie $t$ is important w.r.t. $s$ as it can be reached by many paths from $s$). Our goal is to capture this importance between a pair of nodes (\eg a drug target \textsf{FGFR2} and a tumor suppressor \textsf{TP53} in Figure~\ref{fig:eg-reduced}) in a cancer signaling network in the form of a ``distance'' measure. That is, if a node pair $(s, t)$ has a high PPR value then the ``distance'' should be small. Unfortunately, the PPR values of adjacent node pairs in a graph could vary significantly~\cite{ZY+23}. Consequently, directly using the PPR values as a ``distance'' measure may inject a large variance in their values between the adjacent nodes (\eg adjacent drug targets) and other nodes (\eg oncogenes). To alleviate this challenge, we propose \textit{PEN distance} as follows.

    \begin{definition} \label{def:pen} \textbf{[PEN Distance]}\textit{
        Given a signaling network $G = (V,E)$, the PEN distance between nodes $s \in V$ and $t \in V$ is defined as follows:
\begin{equation}
            P[s,t] = 
            \begin{dcases*}
                0 & if  $s = t$\,, \\[1ex] 
                1-\log(\pi_{d}(s,t)+\epsilon) & if  $s \neq t$\,.
            \end{dcases*}
        \end{equation}           
      where $s \neq t$,  $\pi\textsubscript{d}(s,t)=\pi(s,t)\times{d(s)}$, $d(s)$ is the out-degree of node $s$, and $\epsilon = 1e-5$ where $e$ is the Euler constant.\eat{ If $s = t$ then $ P[s,t]=0$.} \/}
    \end{definition}
In the above definition, $\pi_d(s,t)$ is the \textit{degree-normalized PPR} (DPPR) from $s$ to $t$. According to~\cite{YS+20}, multiplying a node's PPR by its out-degree yields a more precise measure of the strength of connections between nodes. Additionally, employing DPPR helps mitigate the variability of PPR among neighboring nodes. The $\epsilon$ parameter is added to avoid undefined $log(\cdot)$ value when $\pi(s,t)=0$. Intuitively, if $\pi(s,t)$ is large, then $P[s,t]$ tends to be small, \ie well-connected nodes have a closer distance from each other.
 
\begin{example} 
Figure~\ref{fig:pen} illustrates the distributions of PEN distance and PPR values for signaling networks in four cancer types ($\alpha=0.2$). Notably, the PEN distance distribution maintains a consistent shape across all cancer types, while the PPR value distribution shows significant variation among them. As we shall discuss later, the stable shape and range of the PEN distance distribution aid in profiling drug targets across various cancer types.
\EndOfProof
\end{example}
\vspace{-1ex}
\subsection{PEN-Diff}
Next, we introduce the notion of \textit{PEN-diff}, which is based on PEN distance. Let $V_g \subset V_C$ be a set of nodes in a signaling network $G_C$. Consider a node $s \in V_C$. The \textit{average PEN distance} from $s$ to $V_g$, denoted as $\overline{P}[s,V_g]$, is given as follows.
\begin{equation}
           \overline{P}[s,V_g]=\frac{\sum_{i=1}^{|V_g|} P[s,t_i]}{|V_g|}
        \end{equation}
 
 Then, the \textit{single-source PEN-diff} of a node $s$ in $G_C = (V_C, E_C)$, denoted as $P_\Delta[s,V_g]$,  is the difference between the average PEN distance to the the nodes in $V_C-V_g$ and the average PEN distance to the nodes in $V_g$. Formally, it is defined as follows.
 \begin{equation}
           P_\Delta[s,V_g]=\overline{P}[s,V_C-V_g] - \overline{P}[s,V_g] 
        \end{equation}
Given two sets of nodes $V_s$ and $V_g$, the \textit{PEN-diff} of $V_s$ is the average single-source PEN-diff values of the nodes in $V_s$ w.r.t. $V_g$ and rest of the nodes in the network. Formally,
 
\begin{definition} \label{def:diff} \textbf{[PEN-diff]}\textit{
    Given $V_s \subset V_C$ and $V_g \subset V_C$ in a signaling network $G_C=(V_C, E_C)$, the PEN-diff of $V_s$ is defined as follows.
    \begin{equation}
        \overline{P}_\Delta[V_s,V_g] = \frac{\sum_{i=1}^{|V_s|}  P_\Delta[v_i,V_g]}{|V_s|}
    \end{equation}}
\end{definition} 

\begin{algorithm} [t]
\algsetup{
		linenosize=\small
	}

\caption{Target-aware Cancer-specific Signaling Network Construction.}\label{alg: construction}
    \begin{algorithmic}[1]
    	\small
        \REQUIRE Reduced signaling network $G = (V, E)$, the cancer type $C$, set of known target $X_{t}$,  a set of oncogenes $X_{g}$ of $C$, path length threshold $d$
        \ENSURE Target-aware cancer-specific signaling network $G_{C} = (V_{C}, E_{C})$
        \STATE $P_{C} \gets \emptyset$
        \FOR{ $(u,v) \in X_{t},X_{g}$}
         \IF{$\textsc{isPath}(u, v, G)$}
         \FOR{ $\forall p(u, v)$}
            \IF{$\textsc{length}(p(u, v), G) < d$}
                 \STATE $\textsc{add}(p(u, v), P_C)$ 
            \ENDIF
		\ENDFOR           
           \ENDIF
        \ENDFOR
        \STATE $G_{C} \gets \textsc{subgraph}(G,P_C)$
          \RETURN $G_C$
    \end{algorithmic}
\end{algorithm}

 \noindent \textbf{Remark.} Observe that if $ \overline{P}_\Delta[V_s,V_g] > 0$ then the average PEN distance of nodes in $V_s$ and $(V_C-V_g)$ is larger than the average PEN distance with $V_g$. That is, the nodes in $V_s$ are relatively less connected to the nodes in $(V_C-V_g)$ compared to the nodes in $V_g$. In the next subsection, we shall represent a set of oncogenes using $V_g$ and target combinations using $V_s$. Consequently, a positive PEN-diff value for a target combination set indicates that these targets exert relatively less influence on the rest of the network compared to the oncogenes, which is desirable due to off-target effects. Also, note that although in this paper we use $V_g$ to represent oncogenes, \textsc{panacea} is flexible to represent other types of nodes (\eg biomarkers, disease-related nodes) as $V_g$. The framework can be easily applied to all such variations of $V_g$. 
\vspace{0ex}
\subsection{Profiling Drug Target Combinations} \label{sec:algo}
We now present the algorithm to profile the known target combinations in a cancer signaling network by exploiting PEN distance. It is composed of the following four phases. 

\vspace{1ex}\noindent\textbf{Phase 1: Target-aware cancer-specific signaling network construction.} 
Given that cancer is a complex disease involving different genes and proteins for various types (\eg breast, colorectal) and that drug targets differ by cancer type, we first extract a subnetwork from the reduced human signaling network (Table~\ref{tb:summary}) corresponding to a specific cancer type $C$ and the known targets for $C$ (Section~\ref{sec:target}). Algorithm~\ref{alg: construction} outlines the procedure. Given the reduced signaling network $G = (V, E)$, a cancer type $C$, a set of known targets $X_{t}$ for $C$ from the \textit{Anticancer Fund} database~\cite{acf-drugs-db}, a set of oncogenes $X_{g}$ from \textit{COSMIC Cancer Gene Census} database~\cite{sondka2018cosmic}, and a configurable \textit{path length threshold} $d$ ($d=5$ by default), the algorithm searches for paths with length less than $d$ between each pair of a known target $u \in X_t$ and an oncogene $v \in X_g$ in $G$.  The nodes in each such path $p(u, v)$ and their edges are added to a path set $P_C$ and then used to extract the output subnetwork $G_C$ from $G$. That is, the subnetwork encompasses all nodes associated with the oncogenes and targets for $C$, along with the intermediate nodes that link them.

\begin{figure}[t]
    \centering
         \includegraphics[width = \columnwidth]{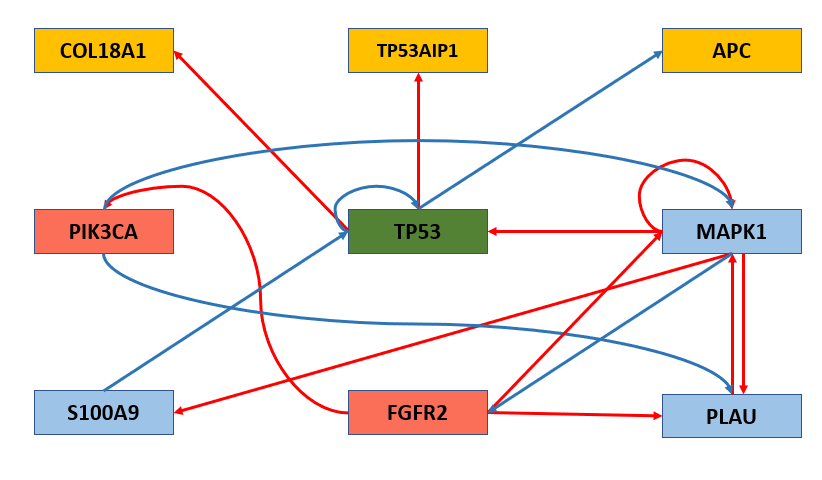}
           \vspace{-2ex}\caption{[Best viewed in color] Target-aware cancer-specific signaling network for breast cancer.}
    \label{fig:breast}
\vspace{0ex}\end{figure}

\begin{table}[t] \scriptsize
    \centering
    \caption{Target-aware cancer-specific signaling networks.}
    \label{tb:summary-cancerntw}
    \begin{tabular}{c|ccc}
    \toprule
   Type & No. of Nodes & No. of Edges & No. of Known Targets\\
    \midrule
    Prostate Cancer & 2,214 & 27,899 & 87\\
    Breast Cancer & 2,560 & 29,986 & 295 \\
    Bladder Cancer & 2,291 & 28,602 & 93 \\
    Colorectal Cancer & 2,467 & 29,717 & 109 \\
    \hline
    \end{tabular}
\vspace{0ex} \end{table}
\begin{example}
Consider the reduced human signaling network in Figure~\ref{fig:eg-reduced}. Assume that we are interested in breast cancer. Then the target-aware, cancer-specific signaling network constructed by Algorithm~\ref{alg: construction} is depicted in Figure~\ref{fig:breast} (partial view). The color coded nodes depict the oncogenes and targets relevant to breast cancer. Table~\ref{tb:summary-cancerntw} reports the features of target-aware, cancer-specific signaling networks for four different cancer types. 
\EndOfProof\end{example}

\noindent\textbf{Phase 2: PEN distance computation.} Given the constructed target-aware cancer-specific signaling network $G_C$, this phase computes the PEN distance between all pairs of nodes in the network by exploiting Definition~\ref{def:pen}. The key step here is the computation of PPR values of pairs of nodes. 
Since this phase incurs a one-time cost for a specific signaling network, we calculate the exact PPR values. It produces the \textit{PEN distance matrix} $\textbf{P}$, which contains the PEN distance values for all pairs of nodes in $G_C$.

\begin{example}
Consider the target-aware, breast cancer-specific signaling network in Figure~\ref{fig:breast}. The PEN distance matrix involving some of the node pairs is shown Table~\ref{tb:all_pdistance}. Observe that 
there are multiple paths from various nodes to \textsf{PIK3CA}, resulting in low PEN distance between these pairs. In contrast, \textsf{TP53AIP1} and the other nodes exhibit high distance, as there are very few (if any) paths linking them to \textsf{TP53AIP1}.
\EndOfProof\end{example} 

\begin{table*}[t] \scriptsize
    \centering
    \caption{Example of PEN distance matrix of selected nodes in the network in Figure~\ref{fig:breast}.}
    \label{tb:all_pdistance}
    \begin{tabular}{c|ccccccccc}
     \toprule
        & \multicolumn{9}{c}{target node} \\
       source node & TP53 & PIK3CA & S100A9 & TP53AIP1 & FGFR2 & APC & COL18A1 & MAPK1 & PLAU\\
       \midrule
         TP53AIP1 & $12.5129$ & $12.5129$ & $12.5129$ & $12.5129$ & $12.5129$ & $12.5129$ & $12.5129$ & $12.5129$ & $12.5129$\\
         FGFR2 & $2.4265$ & $2.6688$ & $3.2374$ & $4.0358$ & $1.3471$ & $4.0358$ & $4.0358$ & $1.4049$ & $1.9927$\\
         APC & $12.5129$ & $12.5129$ & $12.5129$ & $12.5129$ & $12.5129$ & $12.5129$ & $12.5129$ & $12.5129$ & $12.5129$\\
         PIK3CA & $2.763$ & $1.8667$ & $3.5739$ & $4.3722$ & $3.5739$ & $4.3722$ & $4.3722$ & $1.7414$ & $2.3292$\\
         COL18A1 & $12.5129$ & $12.5129$ & $12.5129$ & $12.5129$ & $12.5129$ & $12.5129$ & $12.5129$ & $12.5129$ & $12.5129$\\
         \hline
    \end{tabular}
\vspace{0ex}\end{table*}

\eat{The algorithm of ranking target combinations by pdistance-difference is shown in \ref{alg: ranking-combination}. Given a cancer network $G_{C} = (V_{C}, E_{C})$ and the size of candidate combination $k$, we have $n \choose k$ combinations of size $k$. For each combination, we compute the mean pdistance to cancer genes and non-cancer genes for all elements. Then we take the difference between the two values for each combination and rank them in ascending order.}

\vspace{1ex}\noindent\textbf{Phase 3: PEN-diff computation phase} In this phase, we utilize the PEN distance matrix \textbf{P} to compute the PEN Diff (Definition~\ref{def:diff}) for $k$-node combinations in $G_c$ w.r.t. the oncogenes and non-oncogenes. Algorithm~\ref{alg: ranking-combination} outlines the procedure. Consider the set of oncogenes $X_g$ in $G_c$ and a \textit{combination size parameter} $k$ ($k > 1$). For each $k$-nodes pair of $V_c$, denoted as $H_{k, i}$, we compute the \emph{average} PEN-distance of the nodes in $H_{k, i}$ with the nodes in $X_g$ (Line 3). Similarly, we calculate the average PEN-distance with the nodes in $V_C - X_g$ (Line 4). Finally, the difference between these two values is used to compute the PEN-diff of $H_{k, i}$, which is then stored in the PEN-diff hash table $D$ (Line 5).    

\begin{example} \label{eg:pen-diff}
The PEN-diff values of two 2-node combinations in Figure~\ref{fig:breast} are shown in Table \ref{tb:pen-diff eg}. Notably, the PEN-diff values are positive. The nodes \textsf{FGFR2} and \textsf{PIK3CA} are both targets of the drug \textit{Fulvestrant}, a hormone treatment for advanced breast cancer. This pair has an average PEN distance of $2.4313$ with oncogenes and $3.1732$ with other nodes, indicating that they are, on average, significantly more connected to oncogenes than to the rest of the network. In contrast, the nodes \textsf{FGFR2} and \textsf{TP53AIP1} are not targets of the \textit{same} drug (\ie not a target combination), resulting in a considerably lower PEN-diff value of $0.1611$. 
\EndOfProof
\end{example}

\begin{algorithm}[t]
\algsetup{
		linenosize=\scriptsize
	}

\caption{PEN-diff Computation}\label{alg: ranking-combination}
    \begin{algorithmic}[1]
    	\small
        \REQUIRE Signaling network $G_{C} = (V_{C}, E_{C})$, combination size parameter $k$, PEN distance matrix $\textbf{P}$, the set of $k$-combinations in $V_{C}$ $H_{k}$, set of oncogenes $X_g$ 
        \ENSURE PEN-diff hash table $D$
        \STATE $D \gets \emptyset$
        \FOR{ $k$-combination $H_{k_i}\in H_{k}$}
                \STATE $d_{g} = \textsc{average}(H_{k_i}, \textbf{P}, X_g)$
                \STATE $d_{n} = \textsc{average}(H_{k_i}, \textbf{P}, (V_c - X_g))$
                \STATE $D[H_{k_i}] \gets d_{n} - d_{g}$                 
        \ENDFOR
        \RETURN $D$
    \end{algorithmic}
\end{algorithm}

\begin{algorithm}[t]
\algsetup{
		linenosize=\scriptsize
	}

\caption{Target Combination Profiling}\label{alg:profile}
    \begin{algorithmic}[1]
    	\small
        \REQUIRE PEN Diff hash table $D$, set of known drug targets $X_t$ of a cancer type $C$, number of bucket $N_{bucket}$
        \ENSURE Delta histogram $\mathbb{H}$, target profile thresholds $\delta_{min}$, $\delta_{max}$ 
        \STATE $\delta_{min} \gets 0$
        \STATE $\delta_{max} \gets 0$
        \STATE $R \gets \emptyset$
        \STATE $top \gets [1\%, 10\%, 20\%, 50\%]$
        \STATE $F = \textsc{groupBy}(D, N_{bucket})$
        \STATE $K = \textsc{extractKnownTargetComb}(F, X_{t})$
         \FOR{$m \in top$}
            \STATE $R.\textsc{append}({\frac{\textsc{count}(K>F[m])}{\textsc{len}(K)} \times{100\%}})$
        \ENDFOR 
        \STATE $\mathbb{H} = \textsc{deltaHistogram}(R)$
        \STATE $\delta_{min}, \delta_{max} = \textsc{selectMaxThreshold}(\mathbb{H})$
        \RETURN $\mathbb{H}, \delta_{min}, \delta_{max}$
    \end{algorithmic}
\end{algorithm}

\noindent\textbf{Phase 4: Target combination profiling.}  The PEN-diff hash table $D$ generated in the preceding phase is exploited in this phase to profile known target combinations in $G_C$. Algorithm~\ref{alg:profile} outlines the procedure. Given $D$, a set of known targets $X_t$ for the cancer $C$, and the number of buckets $N_{bucket}$, it first groups the $k$-node combinations in $D$ according to their PEN-diff values (Line 5). Recall that $D$ contains the mapping between a $k$-node combination and its PEN-diff value. Specifically, the \textsc{groupBy} procedure generates equi-width $N_{bucket}$ (by default $N_{bucket}=5$) buckets based on the PEN-diff values and assigns each combination in $D$ to the appropriate bucket $b_i=[r_{min}, r_{max}]$. That is, a node combination $c$ is assigned to a bucket $b_i$ if $D[c] >r_{min}$ and $D[c] \leq r_{max}$. Node combinations within each bucket are sorted by their PEN-diff values in descending order (\ie a lower rank indicates greater similarity in average connectivity or influence between
$V_g$ and rest of the nodes). The known target combinations in $D$, denoted as $K$, are extracted from $F$ (Line 6).  Then the percentage of known $k$-target combinations in a specific bucket is computed by counting the number of known target combinations in the top-$m$ percentage of node combinations by varying $m \in {1, 10, 20, 50}$ (Line 7-9).  This is used to construct the \textit{delta-histogram}  (Line 10). The $X$-axis of the delta histogram represents the buckets, while the $Y$-axis indicates the percentages of known $k$-target combinations in each bucket for different values of $m$. The $[r_{min}, r_{max}]$ values of the bucket with the highest \textit{coverage} are returned as the target profile thresholds $\delta_{min}$ and $\delta_{max}$ (ties are broken arbitrarily). We define the \textit{coverage} of a bucket as the percentage of known target combinations in the top-$50\%$ of its node combinations. These values for a specific cancer-type network and the delta histogram can be used to prioritize $k$-target combinations. It is important to note that we emphasize the percentage of known target combinations among all known target combinations in a bucket, rather than among all node combinations, since the size of the latter can vary across different cancer types and buckets.

\begin{table}[t] \scriptsize
    \caption{Example of PEN-diff computation.}
    \label{tb:pen-diff eg}
    \begin{tabular}{c|ccc}
        \toprule
         2-node pair & $\overline{P}[V_{s},V_{C} - V_{g}]$ & $\overline{P}[V_{s},V_{g}]$ & PEN-Diff\\
         \midrule
       (FGFR2, PIK3CA) & $3.1732$ & $2.4313$ & $0.7419$\\
       (FGFR2, TP53AIP1) & $7.6914$ & $7.5303$ & $0.1611$\\     
       \hline
    \end{tabular}
\vspace{0ex} \end{table}

\begin{figure*}[t]
    \centering
         \includegraphics[width = 0.33\linewidth]{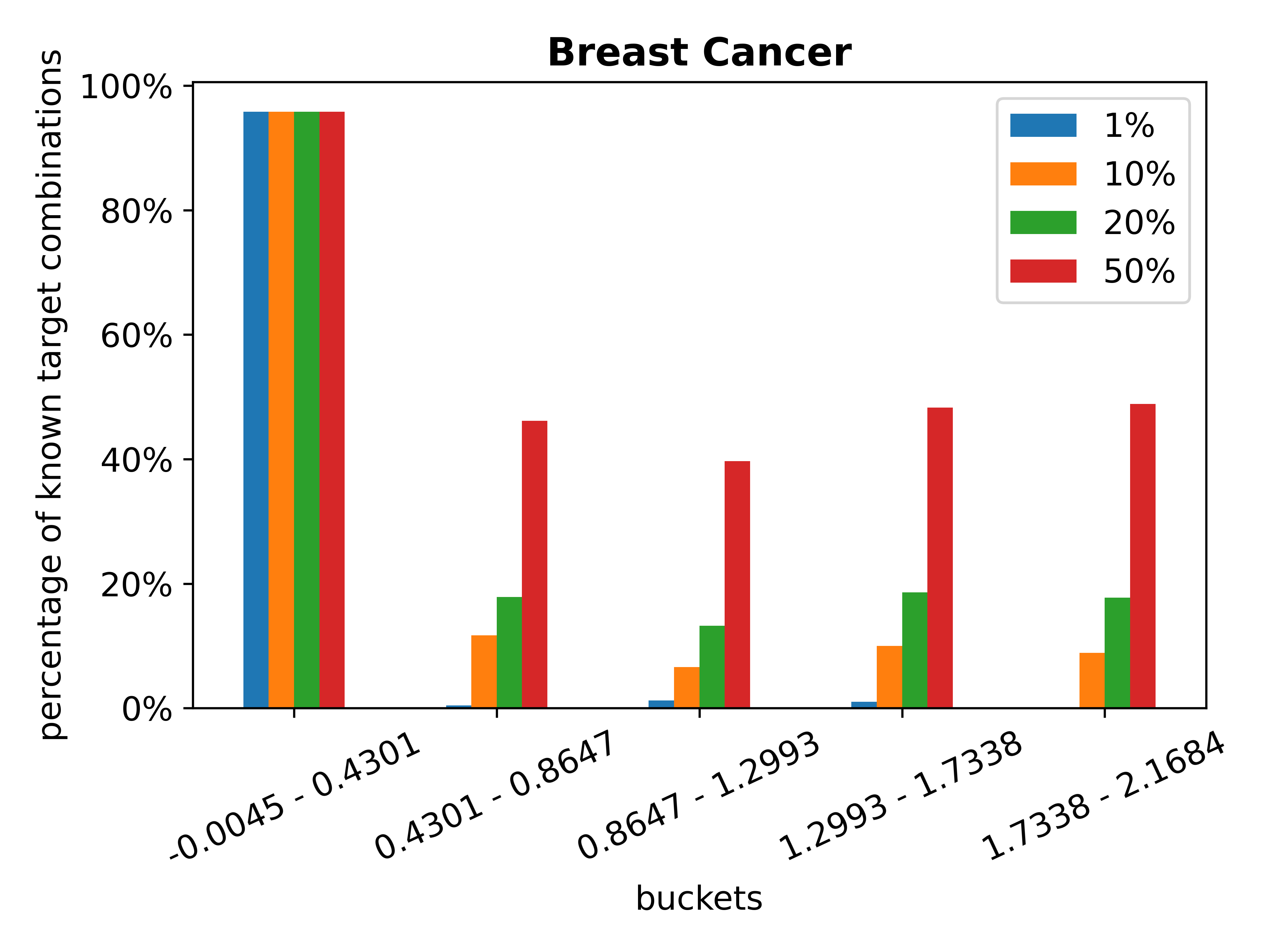}
         \includegraphics[width = 0.33\linewidth]{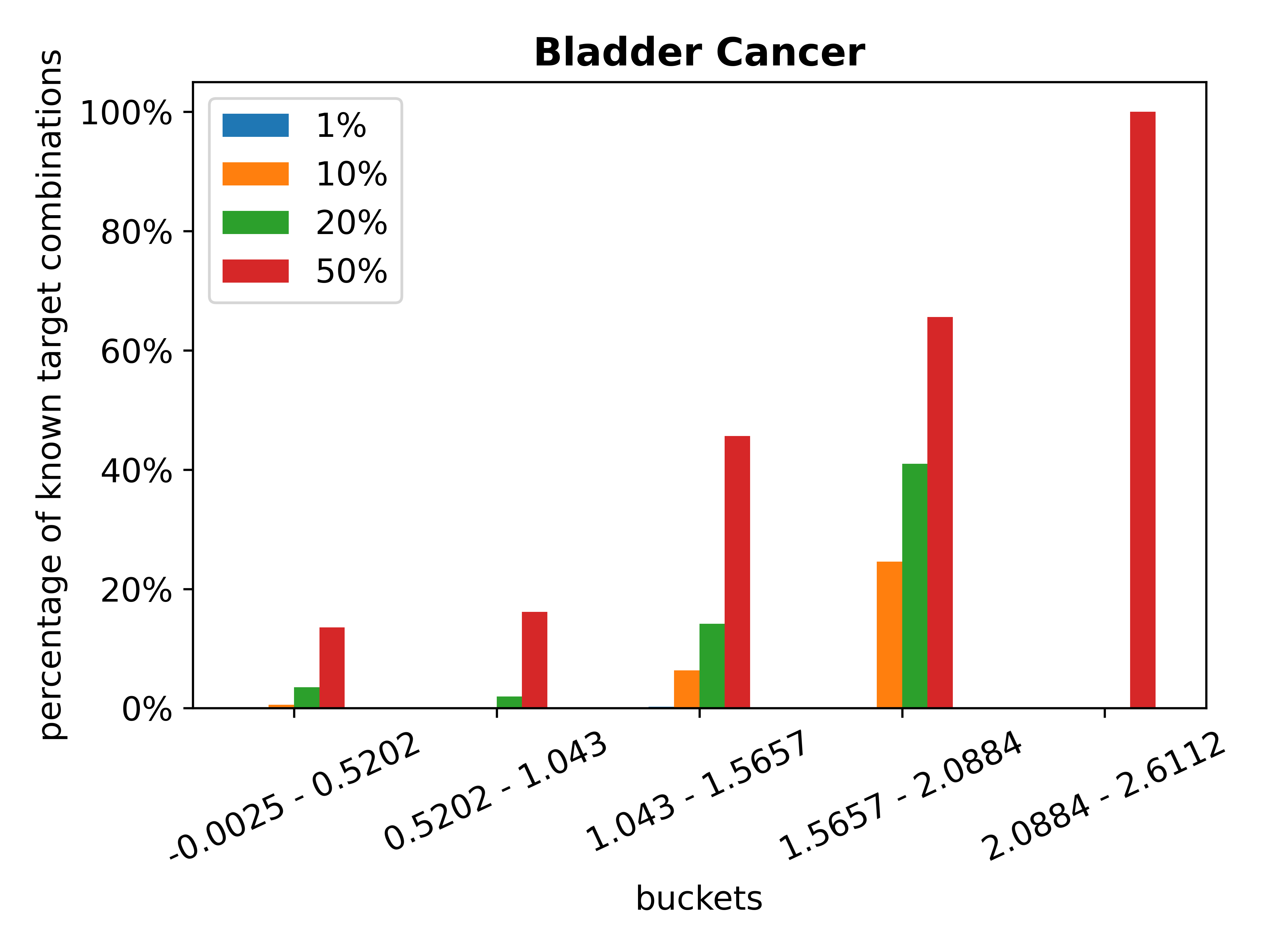} \\
         \includegraphics[width = 0.33\linewidth]{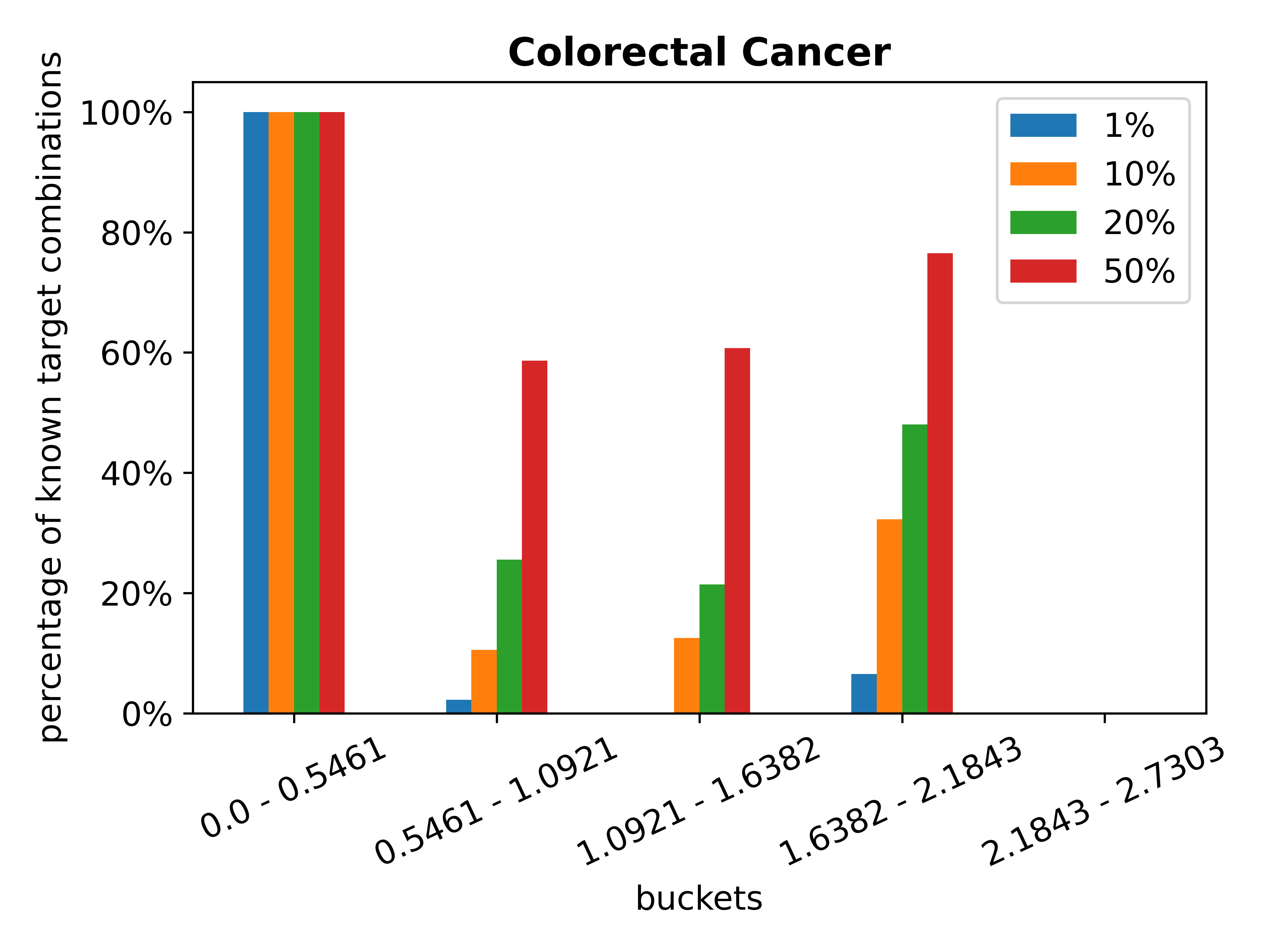}
         \includegraphics[width = 0.33\linewidth]{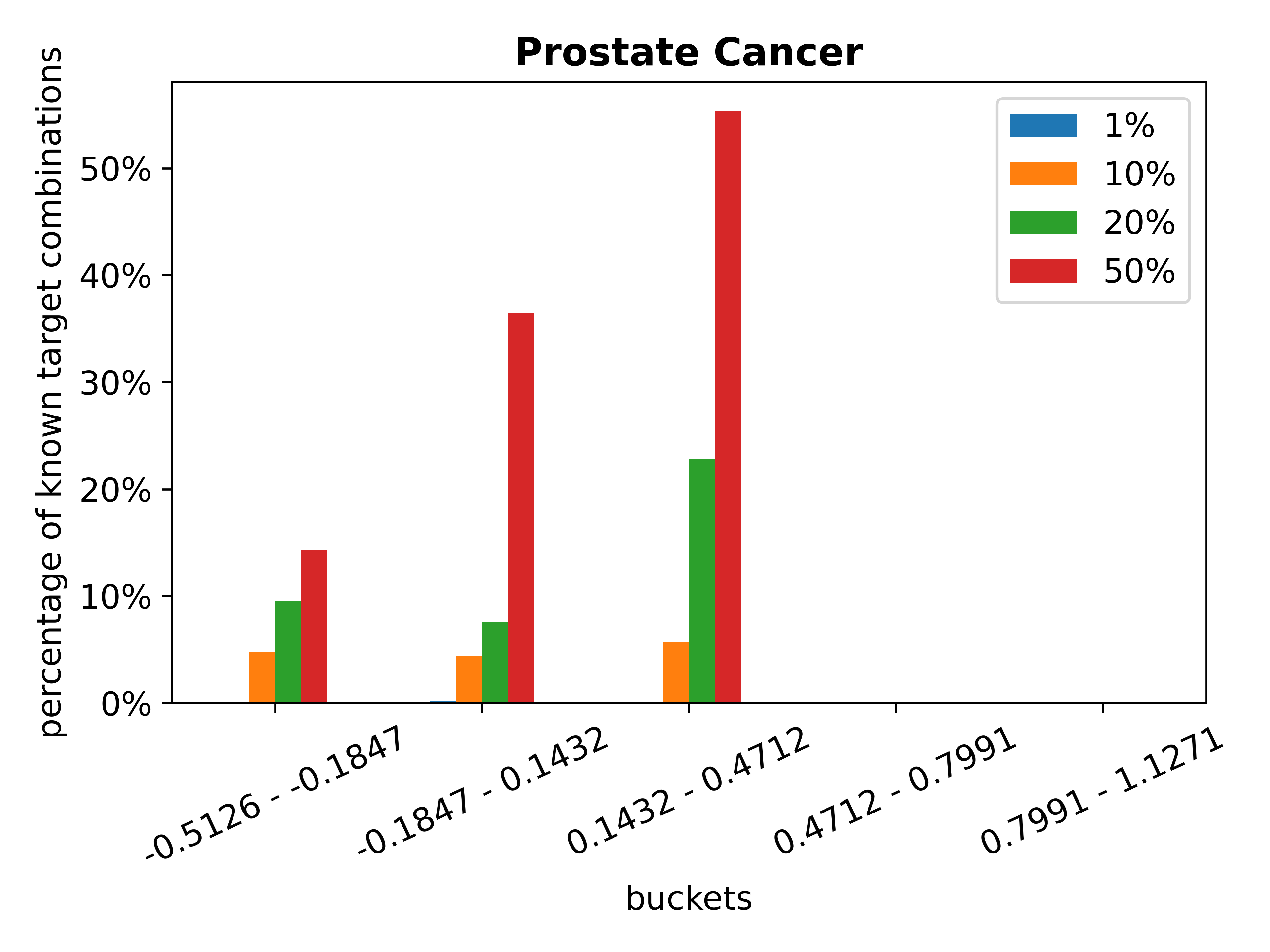}
         \vspace{-2ex}  \caption{[Best viewed in color] Delta histograms of breast, bladder, colorectal, and prostate cancers.}
    \label{fig:delta}
\vspace{0ex}\end{figure*}

\begin{example} \label{eg:delta}
Figure~\ref{fig:delta} (left) depicts the delta histogram of the breast cancer-specific signaling network in Table~\ref{tb:summary-cancerntw}. Observe that majority of the known targets are found in the first bucket. Specifically, the bucket $[-0.0045, 0.4301]$ has the highest coverage of known target combinations for all values of $m$. Hence, these values are returned as $\delta_{min}$ and $\delta_{max}$, respectively, along with the delta histogram. Additionally, most known target combinations fall within the top 1\% of the first bucket, and for the remaining $m$ values, there are no known target combinations beyond those in the top 1\%. As a result, the bars in the first bucket are of equal length.
\EndOfProof
\end{example}

\vspace{-1ex}
\subsection{An Application of Delta Histograms} \label{sec:comb}
The delta histograms characterize the influence of known target combinations on oncogenes in comparison to rest of the nodes in a signaling network. This information is valuable for various downstream applications, including the discovery of novel target combinations. \eat{ Although an exhaustive treatment of the application of delta histograms to these downstream tasks is beyond the scope of the paper,} In this subsection, we will provide an intuitive overview of how delta histograms can be used to efficiently predict novel target combinations in a large cancer signaling network.

First, for a given cancer type, we choose the target pruning threshold and \textit{range constraint} based on the delta histogram. Specifically, we can observe the distribution of the known target combinations and select appropriate bucket(s) for further analysis. Then we select all $k$-node combinations whose PEN-diff values satisfy the range constraint and rank them accordingly. Users can then select the top-$m$ candidates based on their requirements. This prioritized list of target combination candidates can be further examined using a
multi-objective optimization (MOO) algorithm to predict novel target combinations. As we shall see in the next section, this delta histogram-guided approach can significantly reduce the size of the candidate node combination space for MOO compared to other network property-based methods. This, in turn, enables the development of target combination prediction techniques that are well-suited for large, noisy signaling networks.

\begin{example}
Consider the delta histogram of breast cancer in Example~\ref{eg:delta}. Given that the first bucket $[-0.0045, 0.4301]$ has the highest coverage, one may use the range constraint $ \overline{P}_\Delta[V_s,V_g] \leq 0.4301$ to prioritize candidates for target combination discovery. Alternatively, one can examine one or more buckets with fewer known target combinations but higher PEN-diff values to identify potential novel candidates. The delta histogram provides the flexibility to choose bucket(s) based on specific objectives related to target combination discovery. Section~\ref{app:tcp} presents a case study to demonstrate the benefits of delta histograms.
\EndOfProof
\end{example}

\vspace{-1ex}
\section{Performance Study}\label{sec:exp}  
\textsc{panacea} is implemented using Python. We shall now present the key performance results of \textsc{panacea}. All experiments are performed on a 64-bit Windows machine with 12th Gen Intel(R) Core(TM) i7-1250U CPU(1.10 GHz) and 32.0 GB of main memory.
\vspace{-1ex}
\subsection{Experimental Setup}
\textbf{Datasets.} We use the four types of cancer signaling networks in Table~\ref{tb:summary-cancerntw}. We use the sets of oncogenes and known targets as detailed in Section~\ref{sec:back}. We set $\alpha = 0.2$, $k=2$, and $N_{bucket} = 5$ for our experiments. Specifically, we set $k=2$ because most literature on \textit{in silico} target combination discovery primarily focuses on identifying 2 or 3 target combinations for combination therapy~\cite{HC+19,TP21,chua2017synergistic}.  We set  $N_{bucket} = 5$ as it gives a good balance between the number of buckets and the distribution of known target combinations among them. We set $ m \in \{10, 20, 40, 50\}$.

\vspace{1ex}\noindent\textbf{Baselines.} We compare \textsc{panacea} with the following baselines. 

\begin{itemize}
\item \textbf{PPR-diff}: Recall that a key motivation for introducing PEN distance is that we cannot effectively use PPR to profile known target combinations. Therefore, in this baseline, we use PPR (Section~\ref{sec:ppr}) values of the node pairs in $G_C$ to calculate the average difference instead of PEN-diff. 

\item \textbf{Distance-diff}: We leverage network distance between a pair of nodes $s$ and $t$ in $G_C$ (\ie the length of the shortest path from $s$ to $t$) in lieu of PEN distance to compute the average difference.  Note that distance has been exploited by several recent techniques for target/drug combination discovery~\cite{HC+19,CKB19}. Additionally, it serves as the basis for various network centrality measures, including betweenness and closeness centrality. 
\end{itemize}

Note that we do not consider \textit{PDistance}~\cite{ZY+23} as a baseline as it is designed specifically for positioning nodes for network visualization.

The computation of \textit{PPR-diff} and \textit{Distance-diff} is similar to PEN-diff. Given a signaling network $G_{C}$, for a $k$-node combination, we compute the average PPR (resp. distance) to the set to oncogenes $V_{g}$ and rest of the nodes $V_{C} - V_{g}$ in the cancer-specific signaling network. We then find the difference to determine the \textit{PPR-diff} (resp. \textit{Distance-diff}).

\vspace{-1ex}
\subsection{Delta Histograms}
We first report the delta histograms generated by \textsc{panacea} and their characteristics. Figure~\ref{fig:delta} plots the results. We can make the following observations. Firstly, the buckets with the highest coverage vary across different cancers. Specifically, for breast and colorectal cancers, the first bucket shows the maximum coverage, while for bladder and prostate cancers, the buckets $[2.0884, 2.6112]$ and $[0.1432-0.4712]$ have the highest coverage, respectively. This variation is expected given the complexity and heterogeneity of different cancer types. Note that these buckets include 55.28\%-100\% of the known target combinations. Specifically, the buckets with the highest coverage encompass the majority of known target combinations for breast, bladder, and colorectal cancers. Secondly, a significant portion of the known target combinations (with some exceptions for breast cancer) exhibit positive PEN-diff values. This suggests that these known target combinations are more closely connected to the oncogenes than to the other nodes, indicating they exert greater topological influence on the oncogenes compared to the rest. Thirdly, the known target combinations are not always grouped in buckets with high PEN-diff values (i.e., the rightmost buckets) for \emph{every} cancer type. This creates an opportunity to investigate candidate combinations in buckets with elevated PEN-diff values for the discovery of novel target combinations.

\begin{table} \scriptsize
    \centering
    \caption{Characteristics of maximum-coverage buckets.}
    \label{tab:pendiff-delta}
    \begin{tabular}{c|cccc}
    \toprule
    Type & Candidate size & Range constraint & $|b|$ & Coverage\\
    \midrule
    Breast Cancer & $3,275,520$ & $[-0.0045, 0.4301]$ & $1,770$ & $95.81\%$ \\
    Bladder Cancer & $2,623,195$ & $[2.0884, 2.6112]$ & $1,985$ & $100\%$ \\
    Colorectal Cancer & $3,041,811$ & $[0.0, 0.5461]$ & $91$ & $100\%$ \\
    Prostate Cancer & $2,449,791$ & $[0.1432, 0.4712]$ & $741,637$ & $55.28\%$ \\
    \hline
    \end{tabular}
\vspace{0ex}\end{table}

\begin{table}[t]
    \centering \scriptsize
    \vspace{0ex} \caption{Exploration Size Ratio (ESR).}
    \label{tab:max_size}
    \begin{tabular}{c|ccc}
    \toprule
    Cancer Types & $M$ & $|b_{worst, M}|$ & ESR \\
    \midrule
    \multirow{4}*{Breast Cancer} &PEN-diff & $226,049$ & $1.00$ \\
     \cline{2-4} 
     & \textit{PPR-diff} & $539,293$ & $2.39$\\
    \cline{2-4}
        & \textit{Distance-diff} & $862,918$ & $3.82$\\
    \hline
     \multirow{4}*{Bladder Cancer} &PEN-diff & $195,974$ & $1.00$ \\
     \cline{2-4} 
     & \textit{PPR-diff} & $881,789$ & $4.50$\\
     \cline{2-4}
     & \textit{Distance-diff} & $322,523$ & $1.65$\\
    \hline

    \multirow{4}*{Colorectal Cancer} &PEN-diff & $296,818$ & $1.00$ \\
     \cline{2-4} 
     & \textit{PPR-diff} & $542,008$ & $1.83$\\
     \cline{2-4}
     & \textit{Distance-diff} & $649,281$ & $2.19$\\
    \hline

    \multirow{4}*{Prostate Cancer} &PEN-diff & $122,070$ & $1.00$ \\
     \cline{2-4} 
     & \textit{PPR-diff} & $1,131,200$ & $9.27$\\
     \cline{2-4}
     & \textit{Distance-diff} & $432,114$ & $3.54$\\
     \hline
    \end{tabular}
\vspace{0ex}\end{table}

\vspace{-1ex}
\subsection{Comparison of Exploration Space Size}
We hypothesize that the size of delta histogram-guided exploration space is significantly smaller and amenable for downstream analysis. In this set of experiments, we analyze the size of this exploration space and compare it to those generated by the baseline strategies.

\vspace{1ex}\noindent\textbf{Size of the maximum-coverage bucket.} We consider the buckets that have maximum (\ie highest) coverage of known target combinations for the four types of cancer. Table~\ref{tab:pendiff-delta} reports the number of candidate $k$-size combination in the cancer-specific signaling network, along with the range constraints (recall from Section~\ref{sec:comb}) for the buckets with maximum coverage, their size $|b|$ (\ie number of $k$-node combinations), and the coverage of known target combinations. We note that each cancer-specific signaling network contains over two million candidate target combinations, making exhaustive exploration of this space for potential target combinations computationally challenging. 
However, if one opts to explore the bucket with maximum coverage, the exploration space is dramatically reduced by $99.95\%$, $99.92\%$, $99.997\%$ and $69.73\%$ for breast cancer, bladder cancer, colorectal cancer, and prostate cancer, respectively. For instance, if one chooses the bucket $[-0.0045, 0.4301]$ for further analysis  in the breast cancer-specific signaling network, they only need to examine $1,770$ out of a total of $3,275,520$ candidate combinations in the network. This significantly reduces the exploration space and has the potential to enhance the efficiency of downstream tasks.

\begin{figure*}[t]
    \centering
        \includegraphics[width=0.33\linewidth]{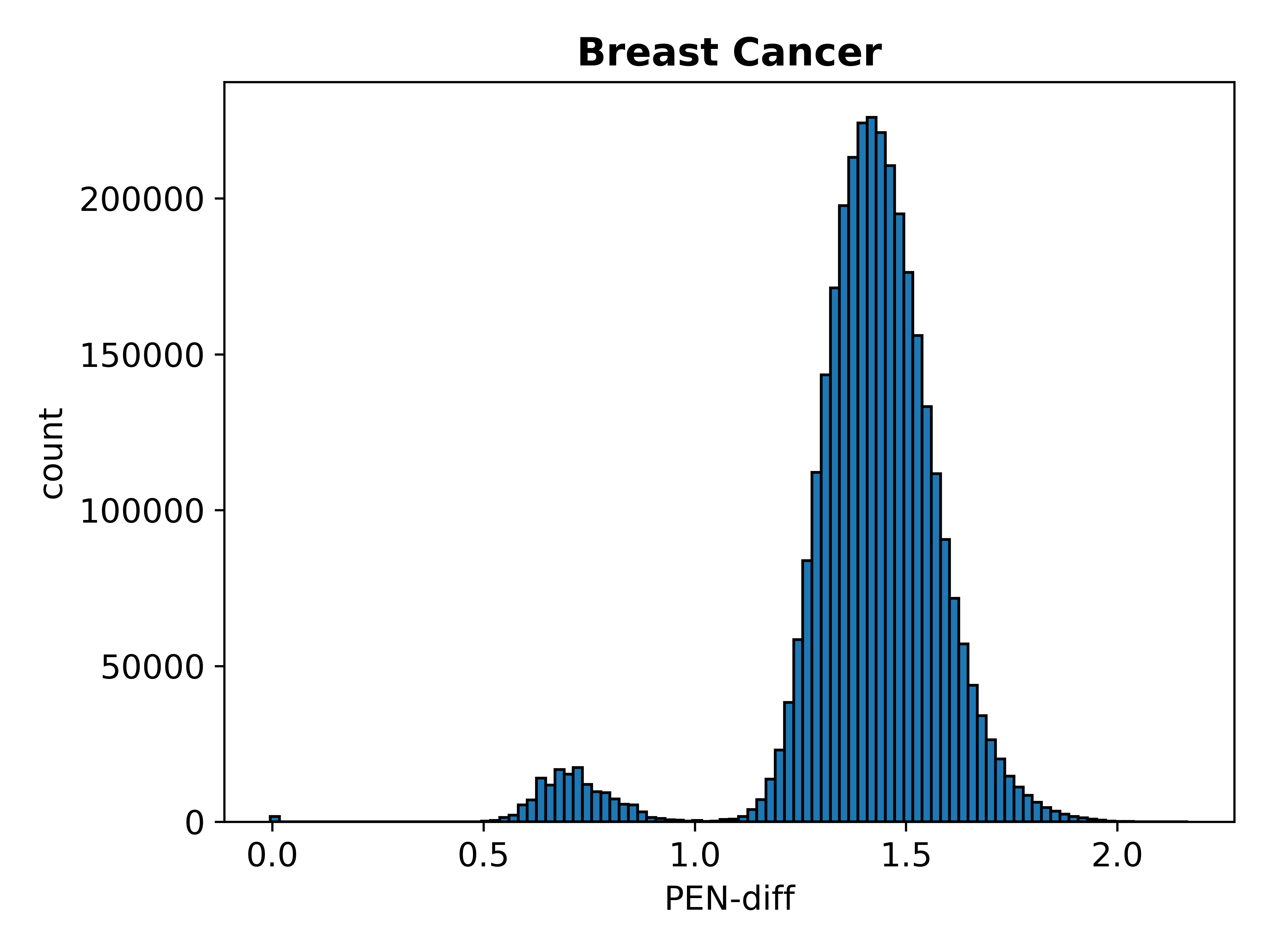}
        \includegraphics[width=0.33\linewidth]{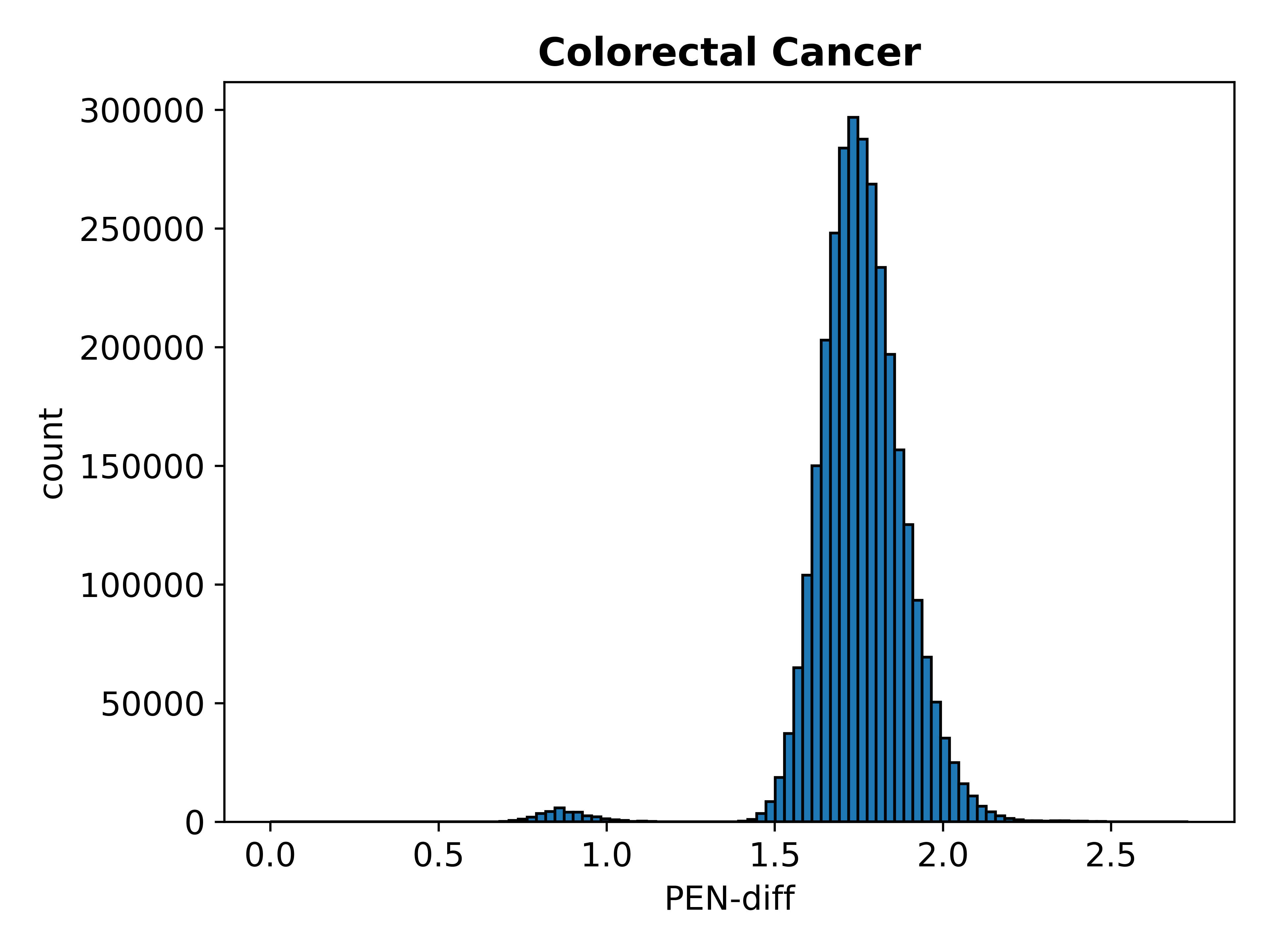}
    \vspace{-3ex}\caption{Distribution of PEN-diff.}
    \label{fig:pendiff-dist}
\vspace{-2ex}\end{figure*}

\begin{figure*}[t]
    \centering
        \includegraphics[width=0.33\linewidth]{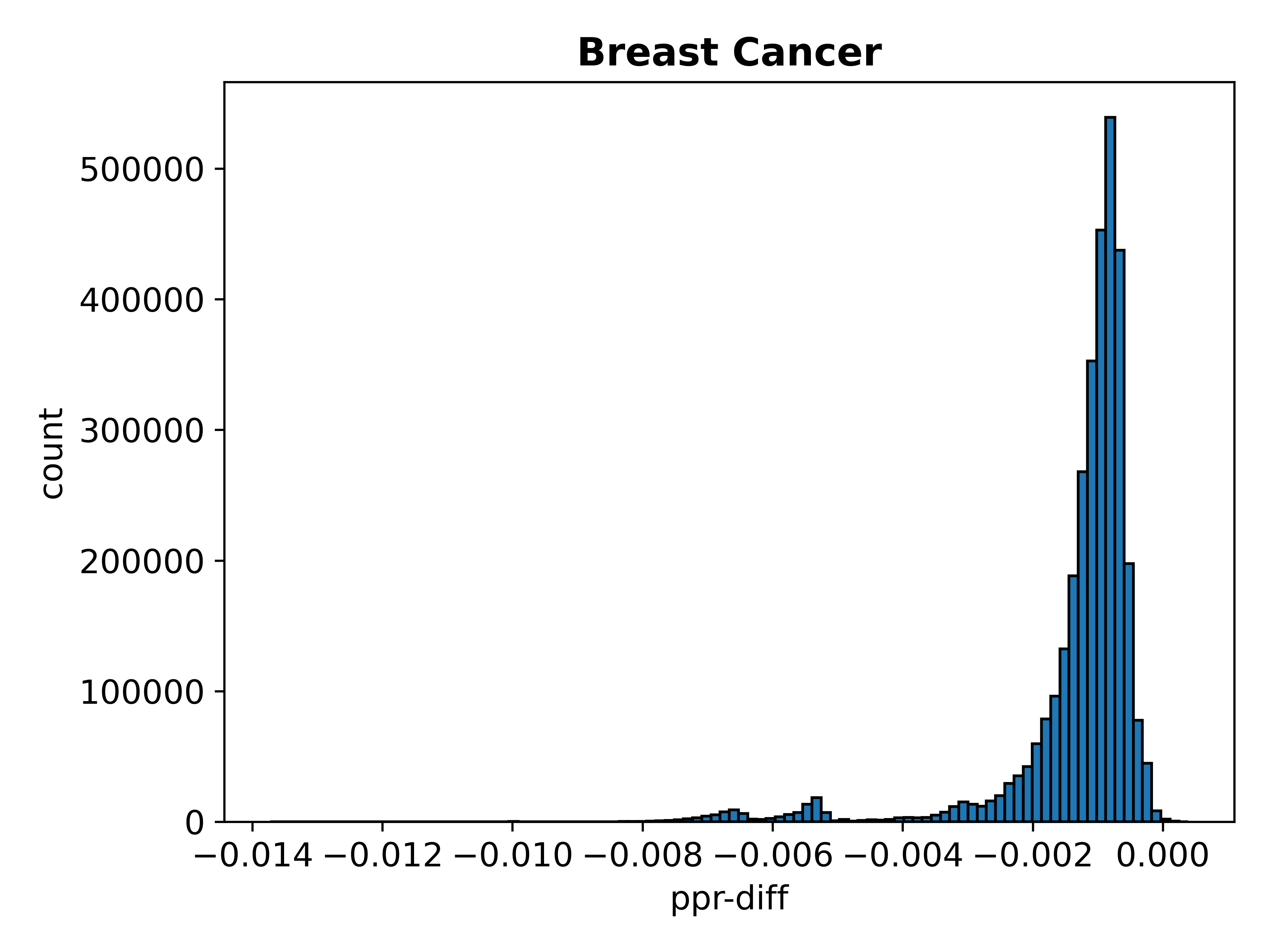}
    \includegraphics[width=0.33\linewidth]{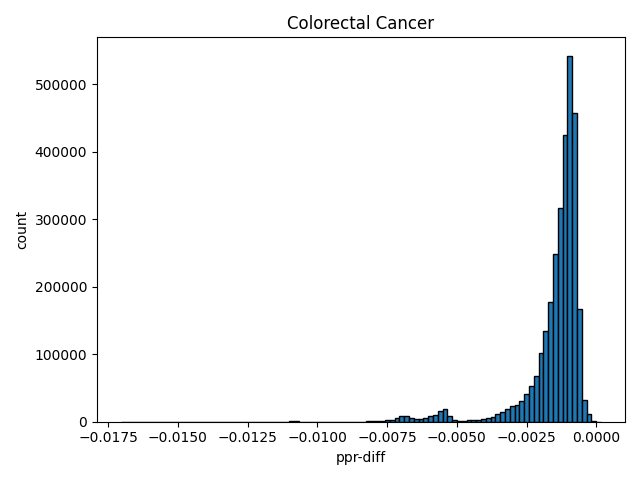}
   \vspace{-3ex} \caption{Distribution of \textit{PPR-diff}.}
    \label{fig:hist-ppr-diff}
\vspace{-3ex}\end{figure*}

\begin{figure*}[t]
    \centering
        \includegraphics[width=0.33\linewidth]{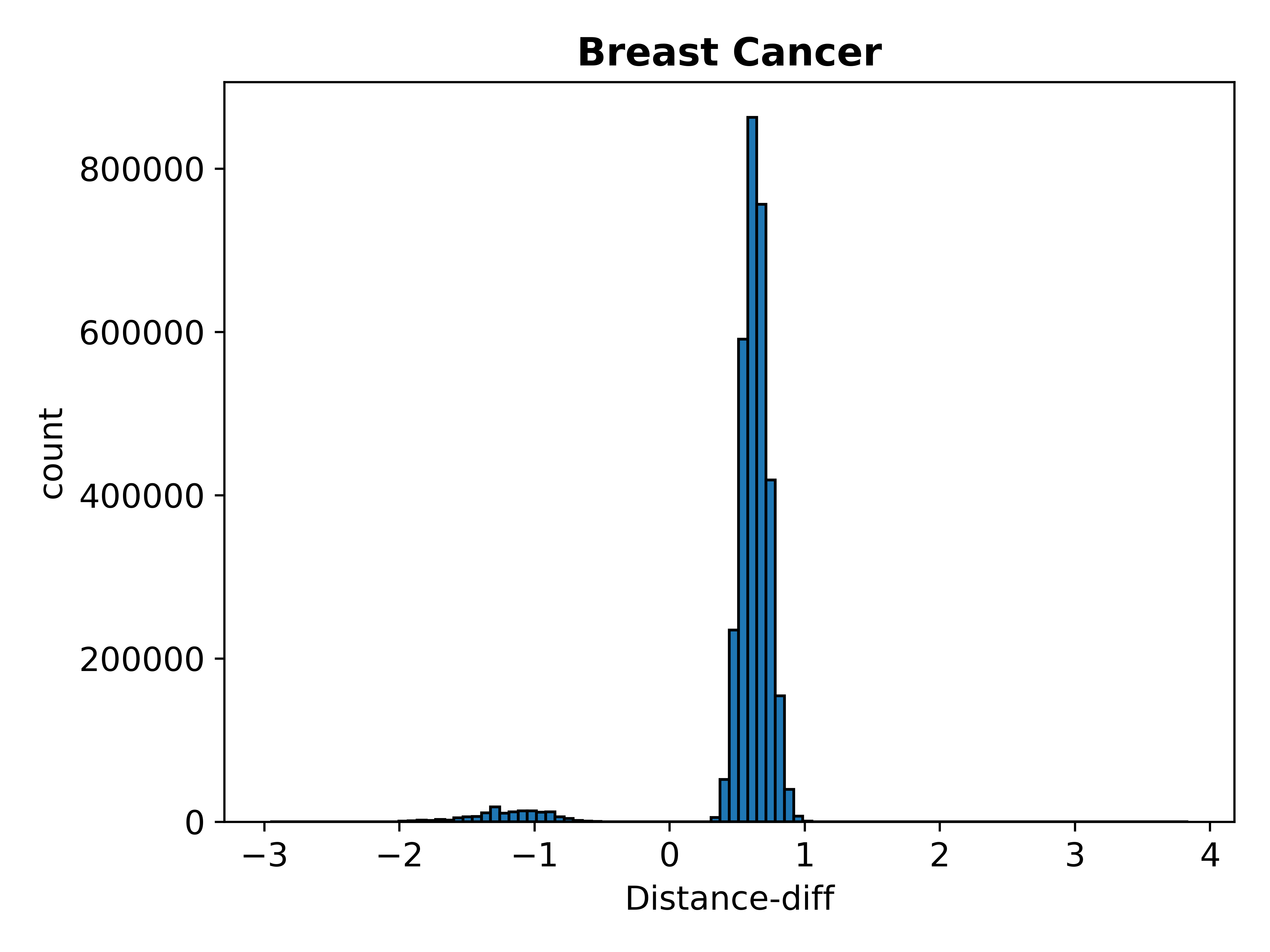}
    \includegraphics[width=0.33\linewidth]{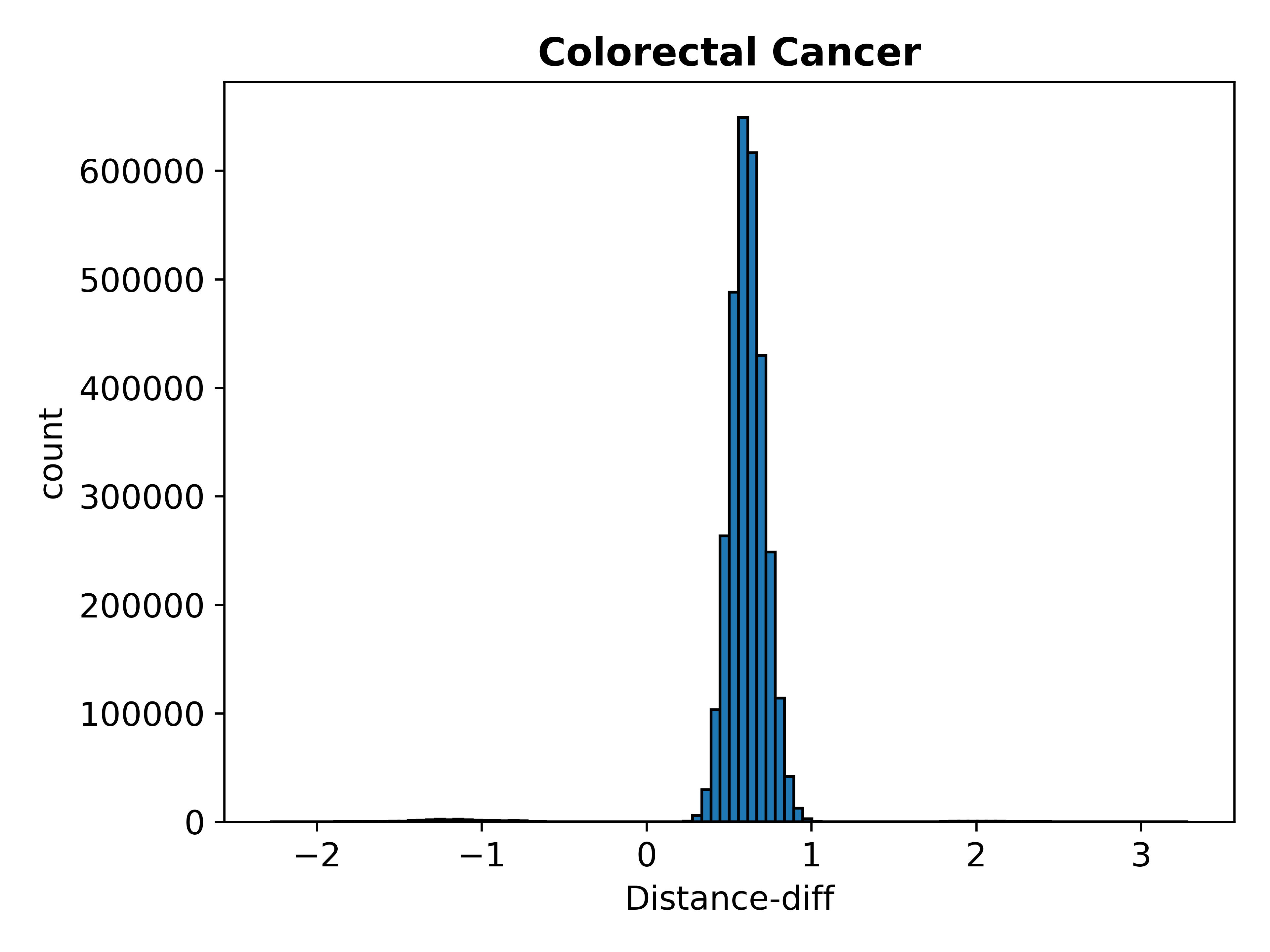}
   \vspace{-3ex} \caption{Distribution of \textit{Distance-diff}.}
    \label{fig:hist-distance-diff}
\vspace{-3ex}\end{figure*}

\vspace{1ex}\noindent\textbf{Comparison of worst-case bucket size.}  The experiment above highlights the benefit of exploring buckets with maximum coverage. Note that one may choose to analyze other buckets as well. Hence, a key question arises: \textit{What is the worst-case size of a bucket in the delta histogram that one may need to explore, and how does it compare to the buckets generated by PPR-diff and Distance-diff?} In this set of experiments, we shed insights into this question. 

\begin{figure*}[t]
    \centering    
        \includegraphics[width=0.33\linewidth]{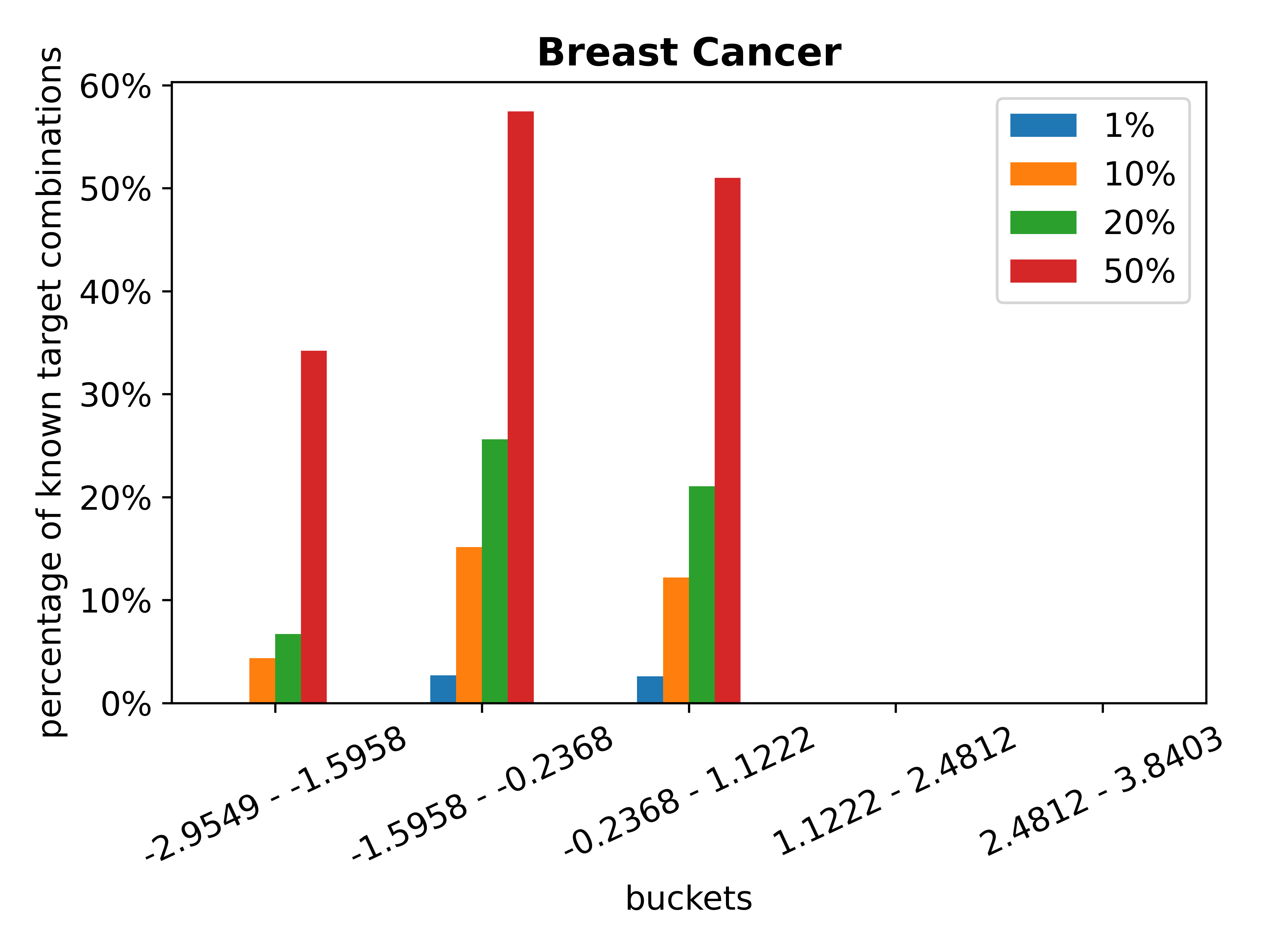}    
        \includegraphics[width=0.33\linewidth]{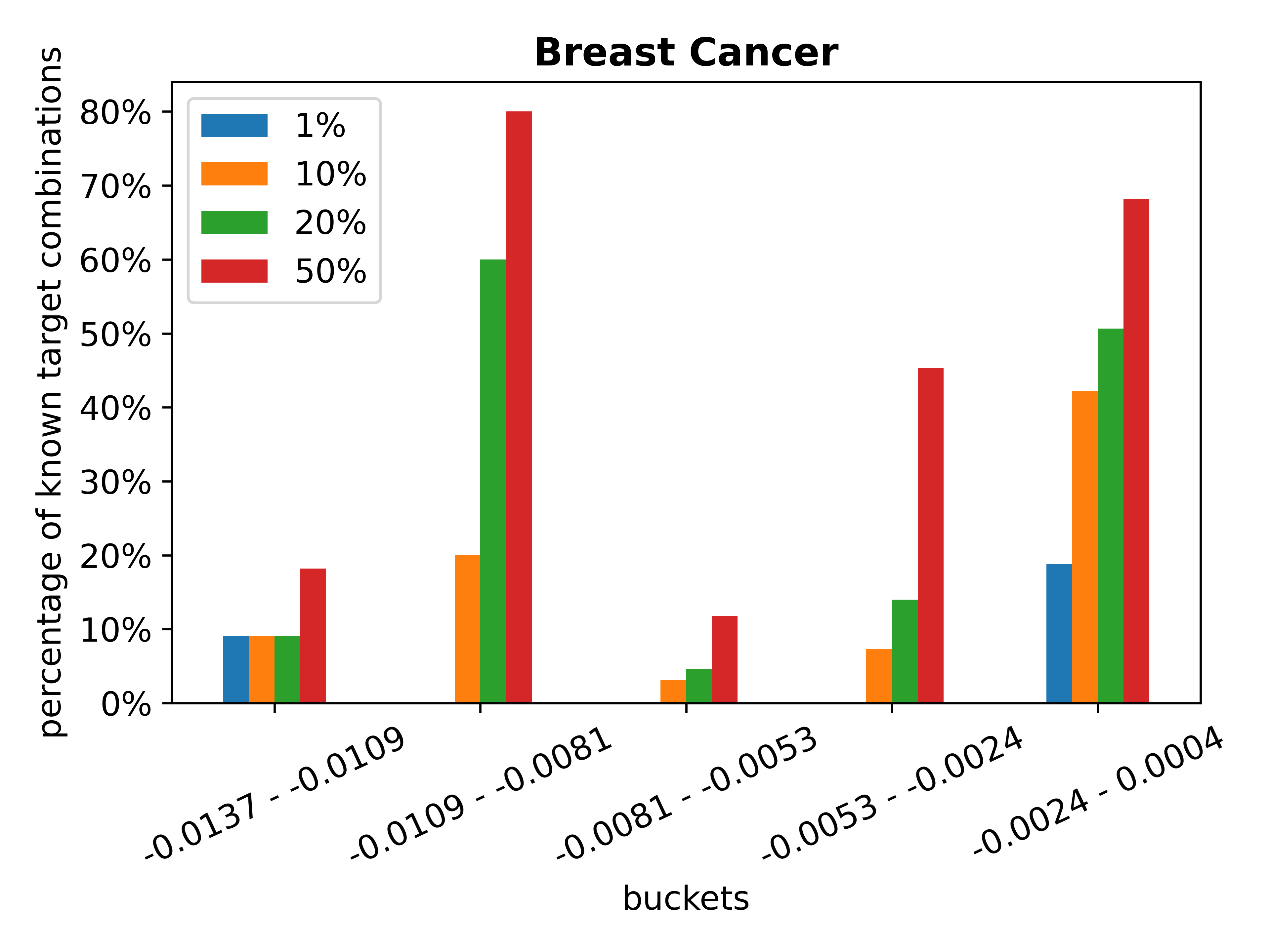}    
    \vspace{-3ex}\caption{[Best viewed in color] Delta histograms generated by baselines for breast cancer: \textit{Dist-diff} (left); \textit{PPR-diff} (right).}
    \label{fig:deltahist-comp}
\vspace{-3ex}
\end{figure*}

We compare the worst-case size of a bucket in the delta histograms of \textsc{panacea}  with the corresponding worst-case bucket sizes in the delta histograms generated by \textit{PPR-diff} and \textit{Distance-diff}, respectively. To this end, we define \textit{exploration size ratio} (ESR) as follows. 
\begin{equation}
ESR_{X} = \frac{|b_{worst, X}|}{|b_{worst, P}|}
\end{equation}
In the above equation, $X \in \{PPR-diff, Distance-diff, \textsc{panacea}\}$, $P$ denotes \textsc{panacea}, and $b_{worst, M}$ is the bucket with the largest number of $k$-node combinations in the delta histogram generated by $M$. Note that $ESR_{\textsc{panacea}} = 1$. The results are shown in Table~\ref{tab:max_size}. Observe that $ESR$  consistently exceeds 1 across all cancer types for both \textit{PPR-diff} and \textit{Distance-diff}. This indicates that, in the worst case, significantly larger candidate combinations must be explored when the delta histograms are generated using network distance or PPR. This occurs because the \textit{PPR-diff} and \textit{Distance-diff} values for most candidate combinations are clustered within a narrow range. Figures~\ref{fig:pendiff-dist}, \ref{fig:hist-ppr-diff}, and \ref{fig:hist-distance-diff} plot the distributions of PEN-diff, \textit{PPR-diff}, and \textit{Distance-diff} values for breast and colorectal cancer. We notice that the distributions of \textit{PPR-diff} and \textit{Distance-diff} have a narrower width compared to PEN-diff. Figure~\ref{fig:deltahist-comp} depicts the delta histograms generated by \textit{Dist-diff} and \textit{PPR-diff} for breast cancer. It is evident that the coverage of the maximum-coverage buckets is lower compared to \textsc{panacea}. The results are qualitatively similar for other cancer types, indicating that these baselines are not effective for profiling known target combinations.

\vspace{-1ex}
\subsection{Can Delta Histograms Aid \emph{In-Silico} Target Combination Prediction?} \label{app:tcp}
In this section, we undertake a case study to demonstrate the benefits of delta histograms generated by \textsc{panacea} for reducing the exploration space for \textit{in-silico} target combination prediction. 

We use the results of Hu \etal~\cite{HC+19} for identifying synergistic \textit{optimal control nodes} (OCN) pairs in breast cancer. The result set is available at \url{https://www.ncbi.nlm.nih.gov/pmc/articles/PMC6522545/bin/41467_2019_10215_MOESM8_ESM.xlsx}. Note that it identifies 63 synergistic OCN pairs as candidates for combination therapy. These pairs involve 28 genes. In our breast cancer signaling network, three of these genes--\textsf{PSENEN}, \textsf{MAML2}, and \textsf{GNG11}--are missing. As a result, we pruned the OCN pairs involving these genes, leading to 35 synergistic OCN pairs. We found that \emph{all} these OCN pairs 
 are confined to just two buckets in the delta histogram for breast cancer, rather than being spread across all five buckets: $[0.8647 - 1.2993]$ and $[1.2993 - 1.7338]$ (Figure~\ref{fig:delta}). Specifically, the PEN-diff values for these OCN pairs fall within $[1.1446 - 1.6398]$. \textit{Therefore, exploring these two buckets instead of the entire breast cancer signaling network can significantly aid in the discovery of all these OCN pairs.} Furthermore,  it is important to note that the PEN-diff values of all these OCN pairs are greater than one. This indicates that these OCN pairs exert relatively less influence on the rest of the network compared to the oncogenes (\ie off-target effects), which is a desirable characteristic, as mentioned earlier.  \textit{Thus, our proposed PEN-diff-based model aligns with the synergistic OCN pairs identified by Hu \etal} Additionally, exploring these two buckets for target combination prediction may uncover novel target pairs with positive PEN-diff values that were not identified by Hu \etal

\vspace{-1ex}
\subsection{Impact of Noise}
Large signaling networks are inherently noisy, and certain areas may lack accurate edge information related to cell signaling. In this experiment, we examine how noise affects delta histograms across various cancer types.  Specifically, we randomly add or remove 1-5\% of the edges in a cancer-specific signaling network and then generate the delta histograms. We present the results for breast and prostate cancer, noting that prostate cancer is the most difficult to profile, with a maximum coverage of 55.28\% (Table~\ref{tab:pendiff-delta}). 

\begin{figure*}[t]
    \centering    
        \includegraphics[width=0.33\linewidth]{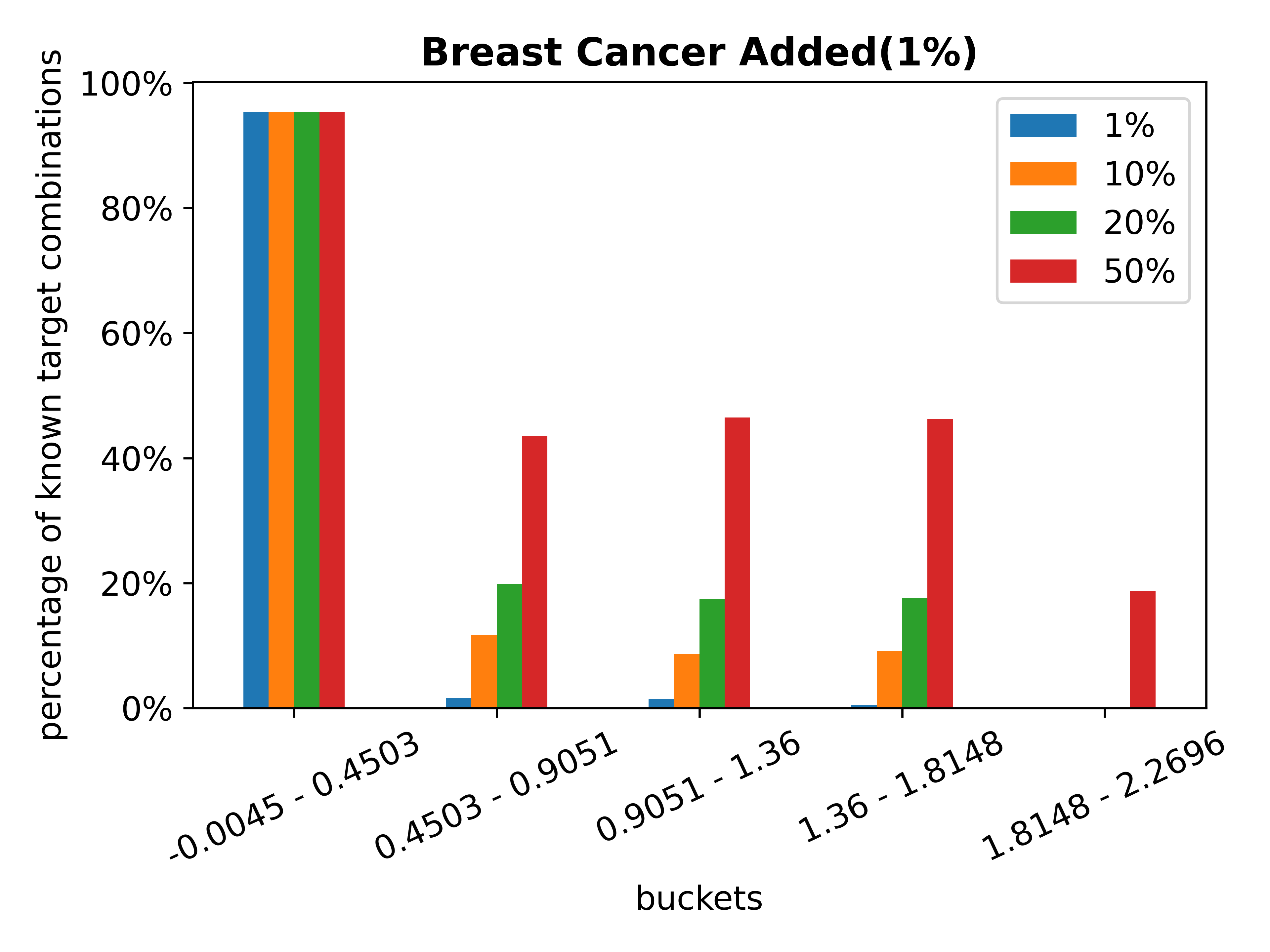}    
        \includegraphics[width=0.33\linewidth]{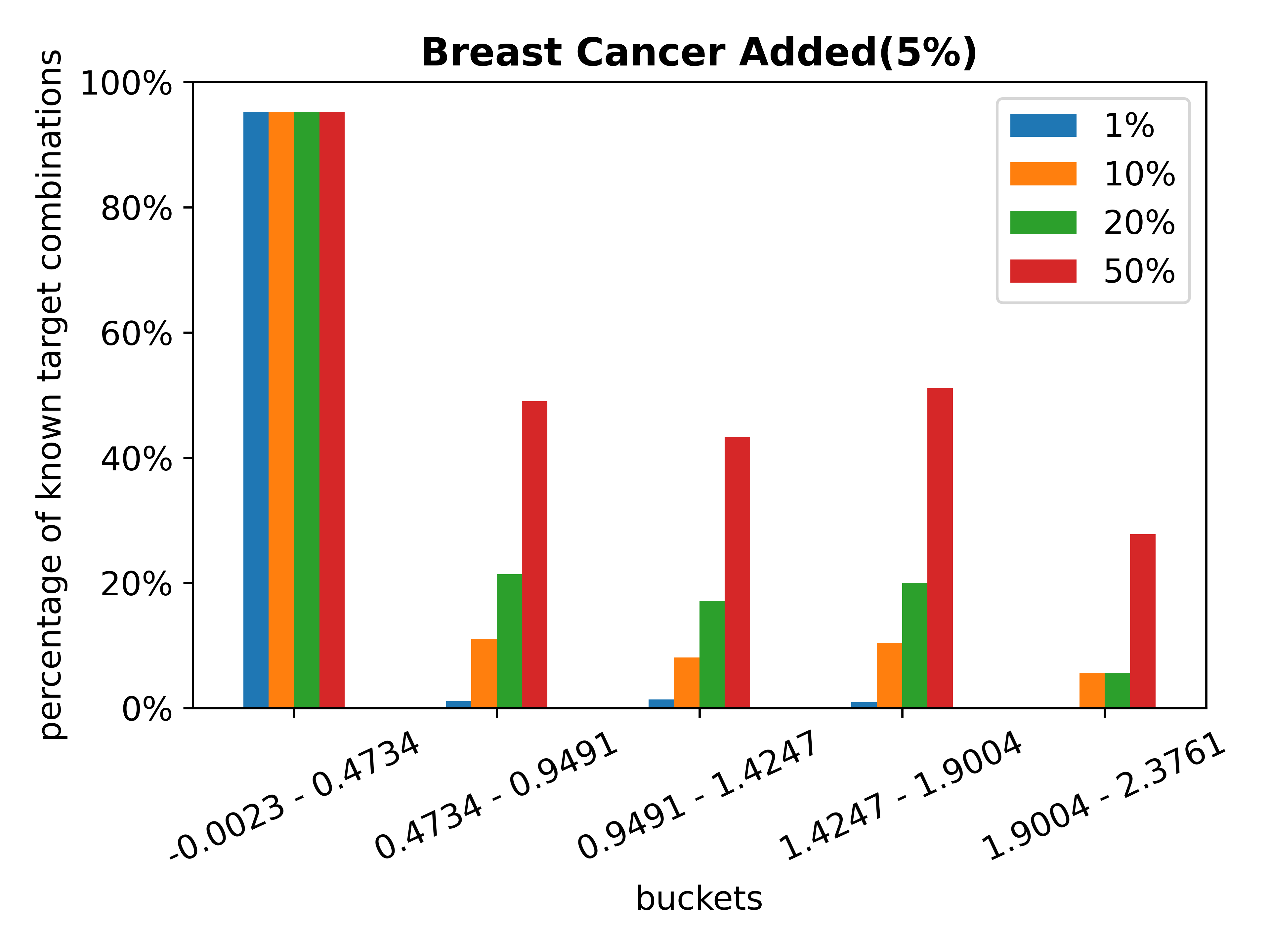}    
    \vspace{-3ex}\caption{[Best viewed in color] Delta histograms for breast cancer in presence of random edge addition: \textit{1\%} (left); \textit{5\%} (right).}
    \label{fig:bc-noise-1}
\vspace{-2ex}
\end{figure*}

\begin{figure*}[t]
    \centering    
        \includegraphics[width=0.33\linewidth]{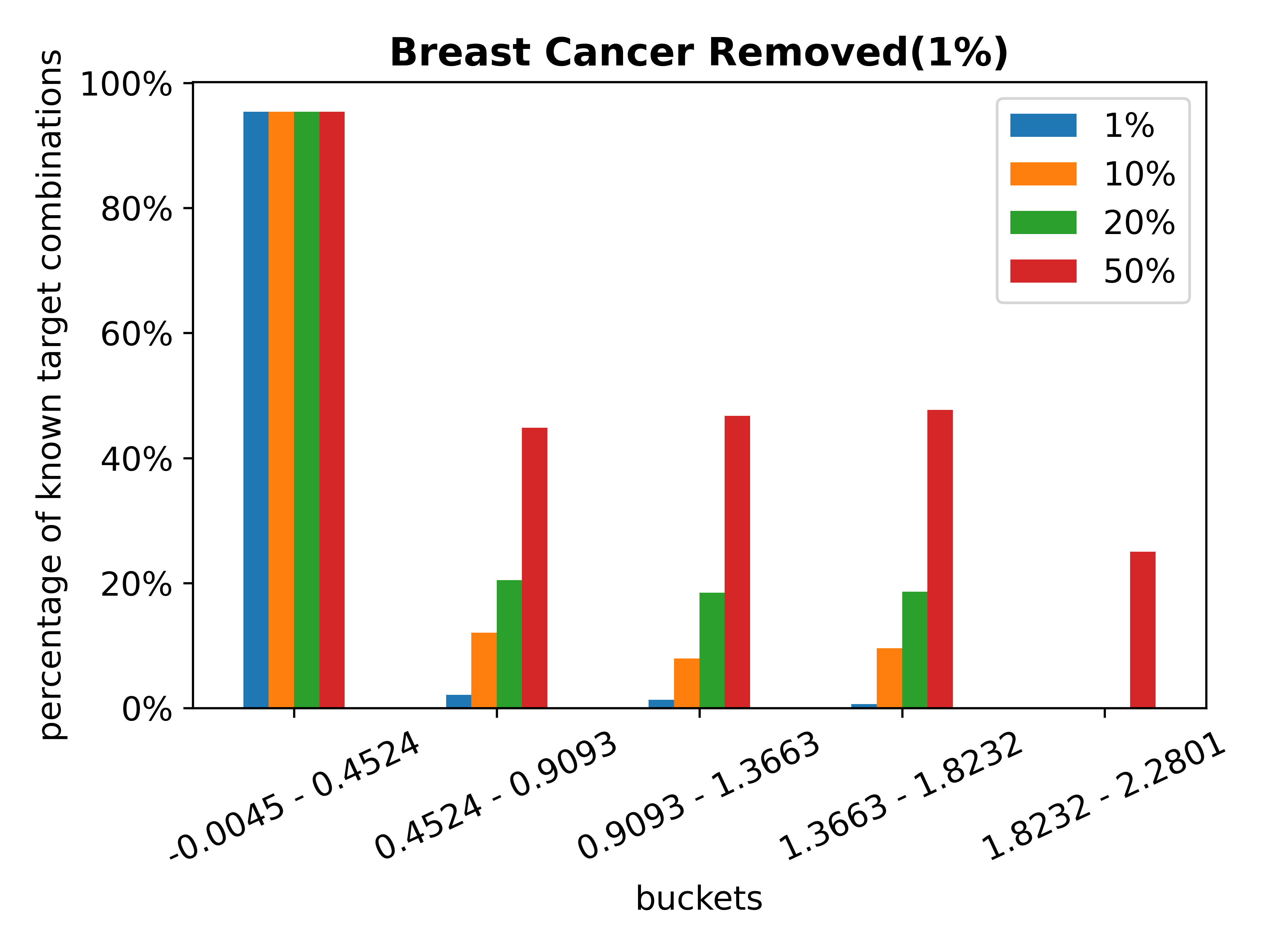}    
        \includegraphics[width=0.33\linewidth]{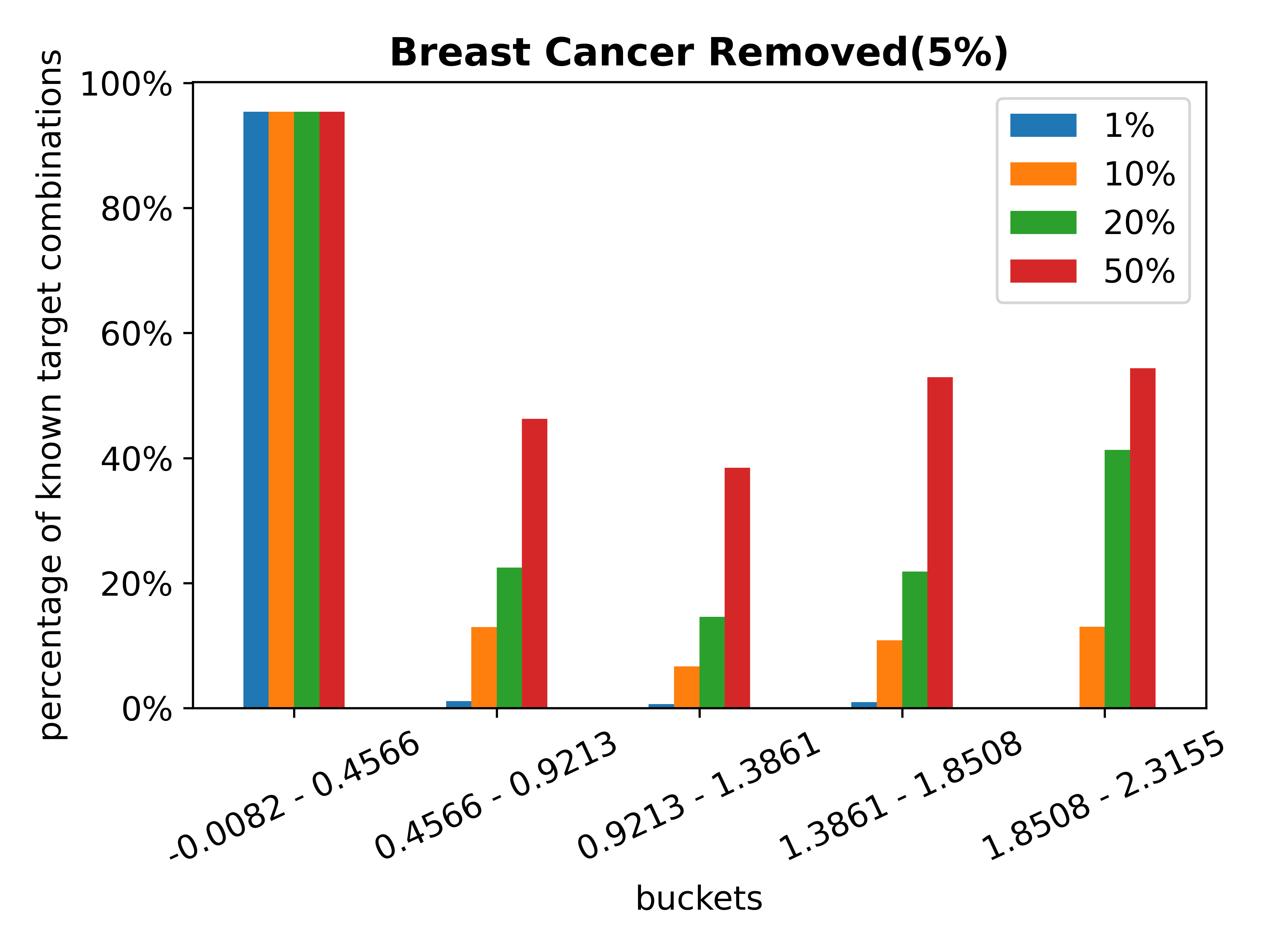}    
    \vspace{-3ex}\caption{[Best viewed in color] Delta histograms for breast cancer in presence of random edge deletion: \textit{1\%} (left); \textit{5\%} (right).}
    \label{fig:bc-noise-2}
\vspace{-2ex}
\end{figure*}

\begin{figure*}[t]
    \centering    
        \includegraphics[width=0.33\linewidth]{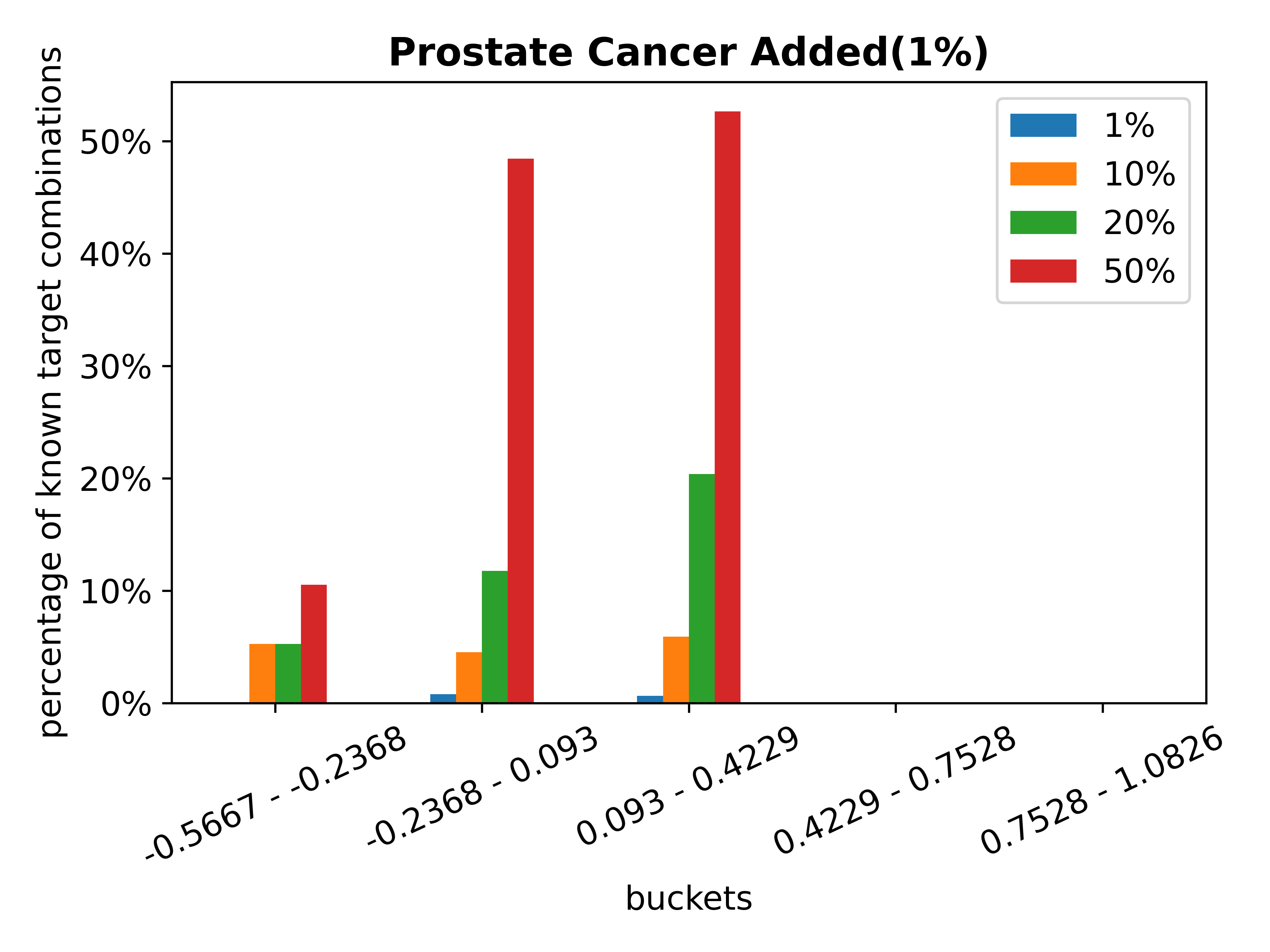}    
        \includegraphics[width=0.33\linewidth]{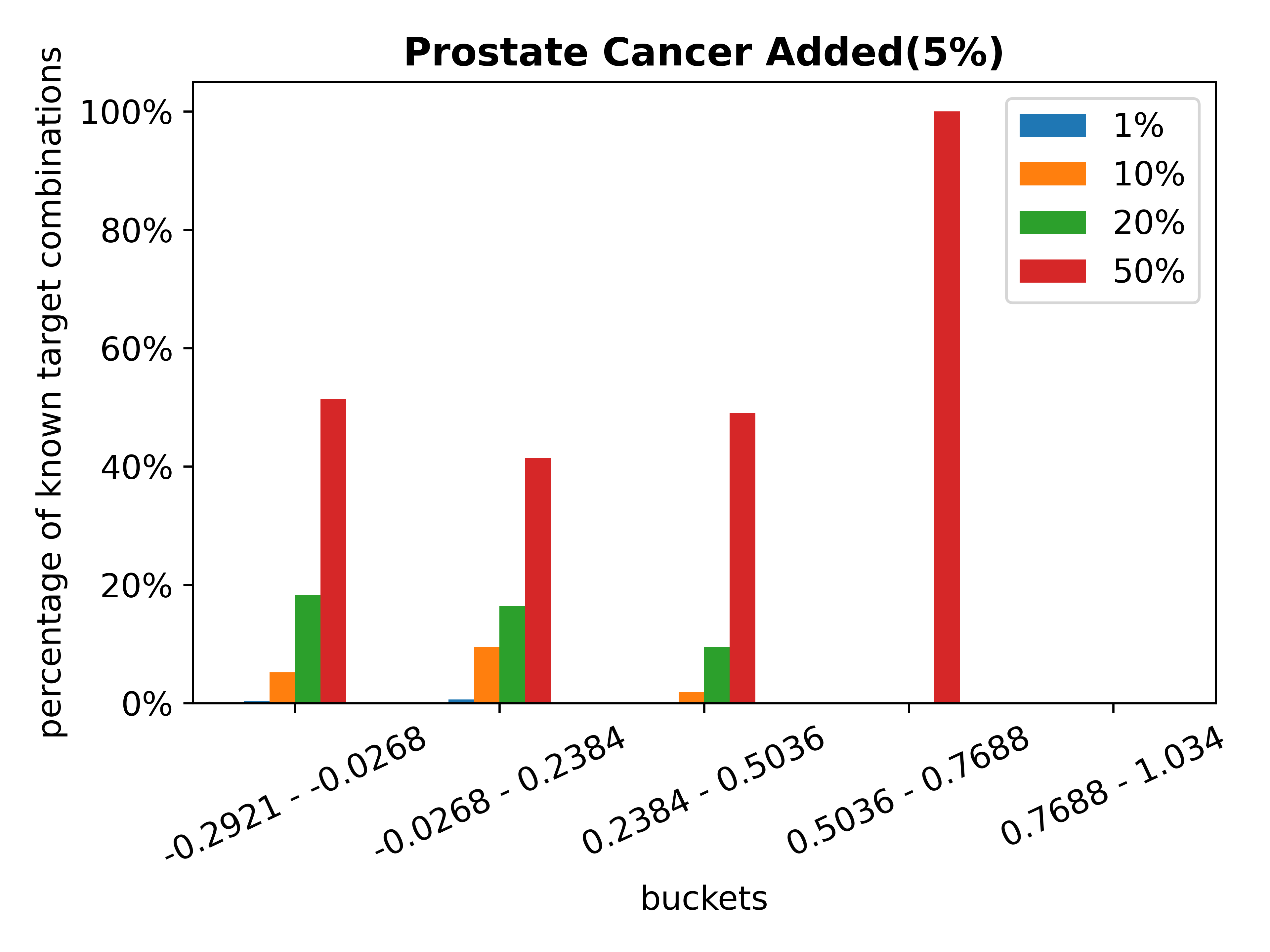}    
    \vspace{-3ex}\caption{[Best viewed in color] Delta histograms for prostate cancer in presence of random edge addition: \textit{1\%} (left); \textit{5\%} (right).}
    \label{fig:pc-noise-1}
\vspace{-2ex}
\end{figure*}

\begin{figure*}[t]
    \centering    
         \includegraphics[width=0.33\linewidth]{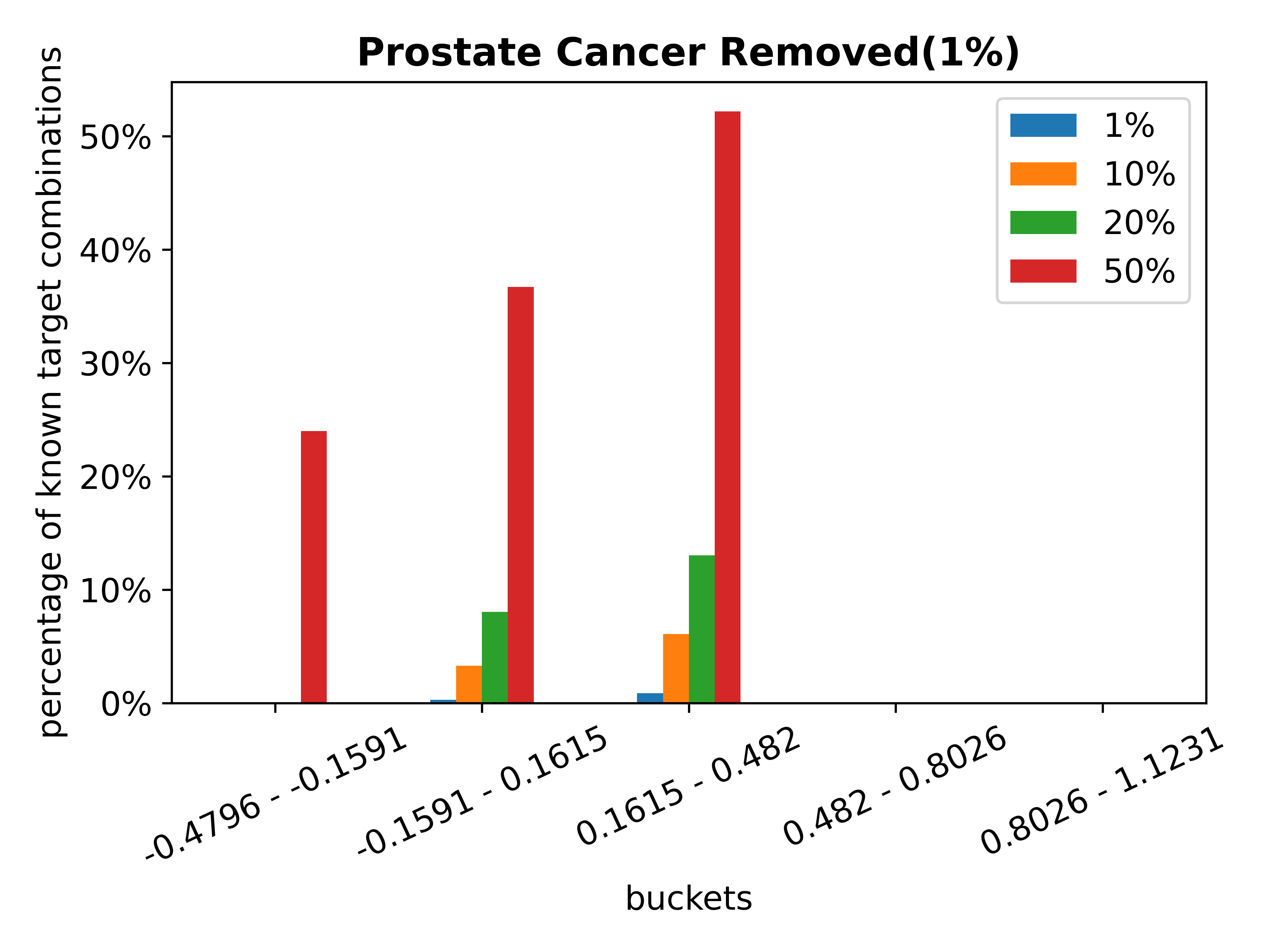}    
        \includegraphics[width=0.33\linewidth]{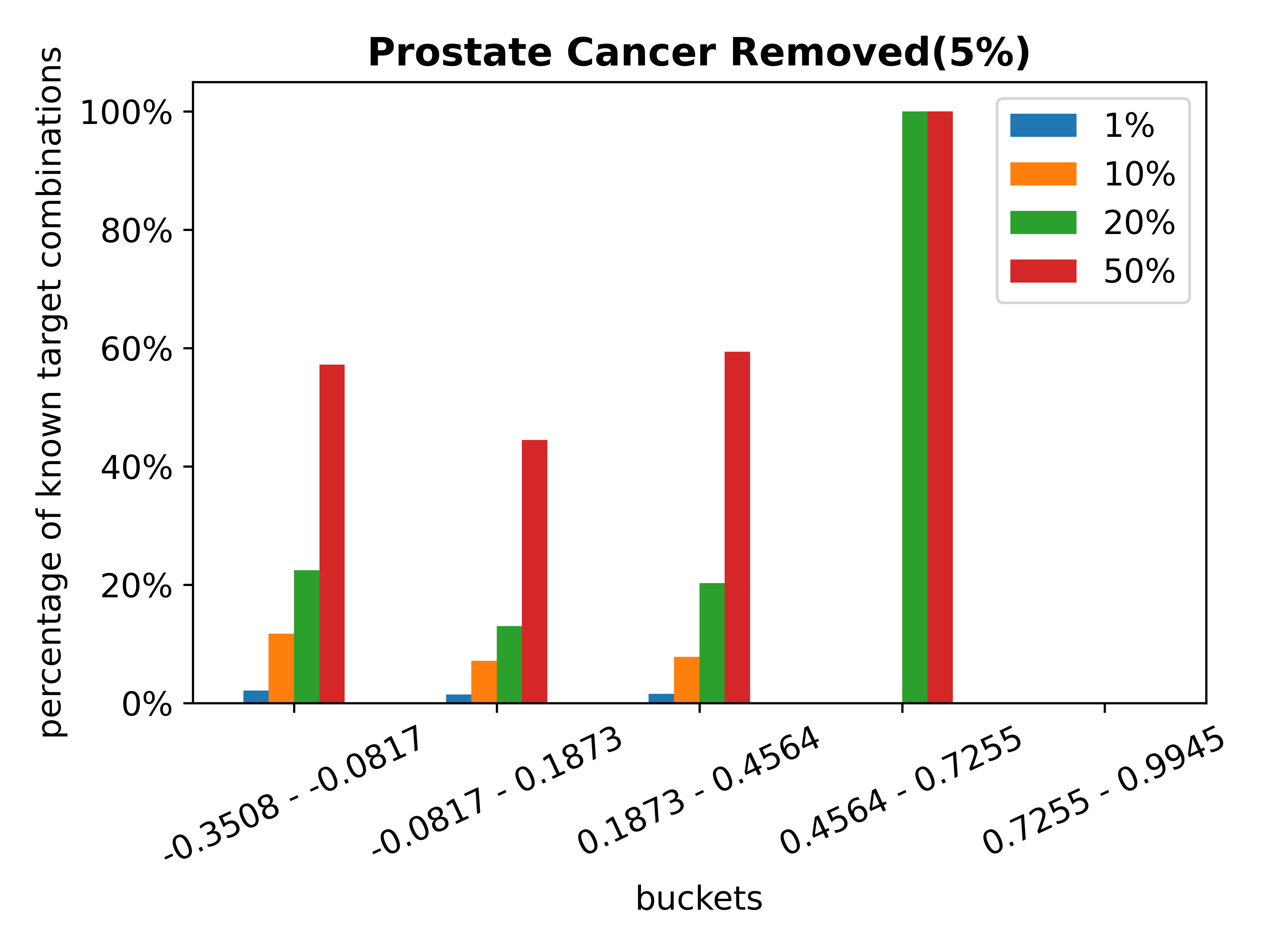}    
    \vspace{-3ex}\caption{[Best viewed in color] Delta histograms for prostate cancer in presence of random edge deletion: \textit{1\%} (left); \textit{5\%} (right).}
    \label{fig:pc-noise-2}
\vspace{-2ex}
\end{figure*}

Figures~\ref{fig:bc-noise-1} and~\ref{fig:bc-noise-2} plot the results for breast cancer. Notably, the shapes of the delta histograms closely resemble those of the original network (Figure~\ref{fig:delta}) with only the $[r_{min}, r_{max}]$ values of the buckets changing. Figures~\ref{fig:pc-noise-1} and~\ref{fig:pc-noise-2} plot the results for prostate cancer. Unlike breast cancer, the shape of the delta histograms for prostate cancer is affected by noise. While the shapes remain similar with a 1\% addition or removal of edges, they change with a 5\% modification, leading to the emergence of an additional bucket. Importantly, \textit{in all cases, there is a dominant bucket with high coverage.} In summary, the shape of the delta histogram varies by cancer type, with some being influenced by the noise in the underlying signaling network, which is expected due to the complexity and heterogeneity of cancer.

\vspace{0ex}
\section{Related Work}\label{sec:rel} 
Significant research has been dedicated to data profiling~\cite{AG+18}, with a primary emphasis on creating various techniques to efficiently uncover statistics and metadata (\eg dependencies, correlations) from data, which can be applied to areas like data cleaning and query optimization. When it comes to network data, profiling efforts have mainly concentrated on analyzing the distributions of metrics such as degree, path length, and different centrality measures, along with summarizing graph structures~\cite{MKC23,BS16}. To the best of our knowledge, none of these studies have specifically addressed the profiling of disease-related signaling networks.

There is extensive research on efficient computation of PPR with quality guarantees~\cite{YW+24}. In this context, we investigate how two PPR-based measures, PEN distance and PEN-diff, can be utilized to profile known target combinations in cancer signaling networks. Zhang \etal~\cite{ZY+23} introduced a PPR-based distance measure called \textit{PDistance}, designed to enhance graph visualization by strategically positioning nodes. While PEN distance also uses degree-normalized PPR, it has a different definition and application than PDistance. PPR has also been used for analyzing protein-protein interaction (PPI) networks~\cite{IG11}. 

Techniques such as random walks, random walks with restart (\eg PPR), and diffusion kernels have been used for network propagation aimed at tasks like function prediction, gene prioritization, module detection, patient stratification~\cite{CI+17}. However, these methods typically yield scoring vectors or similarity matrices rather than delta histograms, and they do not focus on profiling known target combinations. More recently, Pak \etal~\cite{Pak23} developed NetGP to predict drug responses by using drug target information to compute gene perturbation scores through network perturbation methods. In contrast, our research centers on profiling known target combinations in cancer signaling networks using a PPR-based approach and generates delta histograms.

Cheng \etal~\cite{CKB19} investigated network-based relationship between drug-target modules and disease modules in PPI networks, specifically using a distance-based separation measure to assess the proximity of targets. In contrast, our approach employs PEN distance to examine the relationships between a set of drug targets, oncogenes, and the remaining nodes in the network.

Ulgen \etal~\cite{UOS23} developed a framework for personalized drug prioritization in PPI networks using random walk with restart. However, their approach differs from ours in several significant ways. Firstly, it does not focus on profiling known drug target combinations, which is a unique aspect of our study; instead, it prioritizes drugs. Secondly, like many existing methods, it relies on ``shortest distance'' (network distance) instead of PEN distance. Thirdly, they use PPI networks rather than signaling networks. Finally, the algorithms differ primarily because they address a different problem than we do.

Lastly, there is a growing body of research aimed at discovering synergistic target combinations \textit{in silico}~\cite{Chua2012,chua2017synergistic,timma,HC+19}. However, these methods struggle to scale to signaling networks with thousands of nodes. There is also considerable research in synergistic drug combination prediction using various drug features and/or cell line features~\cite{JK+18,CKB19,Remzi19,LT+20,SS+18,Sun15}. As noted in Section 1, our approach, \textsc{panacea}, complements these efforts by generating a delta histogram that can help guide the selection of candidate $k$-node combinations in a cancer signaling network.
 
\vspace{0ex}
\section{Conclusions \& Future Work}\label{sec:conc} 
This paper integrates data profiling with cancer signaling networks and drug targets by introducing the novel influence-driven target combination profiling problem and presenting \textsc{panacea} as a solution for large cancer signaling networks. It utilizes two innovative personalized PageRank-based measures, PEN distance and PEN-diff, to summarize the distribution of known targets in relation to their influence on cancer-mutated genes and other nodes in the network through delta histograms. Experimental results show that the PEN-diff-based profiling outperforms several alternative methods.

This paper represents the first attempt to profile known target combinations in disease-related signaling networks using Personalized PageRank (PPR), opening up several intriguing challenges. The size of the exploration space in large cancer-specific signaling networks has been a longstanding challenge, which has limited existing target (or drug) combination techniques. These methods often overlook the topological profiles of known combinations and must operate on small signaling networks or subgraphs. Delta histograms could help guide and narrow the exploration space for these tasks, making the design of delta histogram-guided target combination prediction solutions a crucial area for future research.

Additionally, a key aspect of \textsc{panacea} is that it does not rely on complete signaling network models or specific genomic and proteomic information, instead adopting a topological approach based on the assumption that large signaling networks often lack comprehensive mathematical models and patient data. When partial data is available for certain regions of the network, it presents an interesting challenge to profile known targets by exploiting them.

Last but not least, the concept of delta histogram-based profiling of cancer signaling networks can be applied to other ``network diseases'' (\eg diabetes). This approach allows us to characterize approved target combinations associated with these diseases and leverage them for predicting new target combinations. Consequently, this work opens up a novel avenue for addressing the multi-target selection problem in various complex diseases.

\begin{acks}
The authors are partially supported by the AcRF Tier-2 Grant MOE2019-T2-1-029.
\end{acks}

\newpage
\appendix

\section{Target Combination Discovery in Drug Discovery Pipeline} \label{app:tcd}
The target-based drug discovery pipeline for combination therapy consists of three key steps: the \textit{discovery phase}, the \textit{preclinical phase}, and the \textit{clinical phase}. In the \textit{discovery phase}, potential targets are identified, screened, and validated, followed by the identification and optimization of lead compounds for these targets. During the \textit{preclinical phase}, these optimized compounds are tested in animal models to assess their efficacy and safety before moving on to human trials in the \textit{clinical phase}. This pipeline typically spans 12--15 years and costs as much as 1 billion USD to bring a single drug to market~\cite{CP13}.

Unfortunately, the drug discovery pipeline has not yielded many successful new drugs. A significant issue is that novel targets identified in the discovery phase tend to have a low success rate~\cite{Dodd05}. Only a small fraction of these targets advance to preclinical studies~\cite{Vogel07}. Most combination therapy drugs were originally developed as effective single agents and were later combined for clinical use~\cite{SS+18}. Since drug effects vary with dosage, multiple doses must be tested for combinations, leading to an exponential increase in possible combinations. While high-throughput screening (HTS) technology facilitates the testing of drug pairs at various doses, the combinatorial explosion makes it impractical to exhaustively measure combinations involving more than two drugs. Therefore, it is essential to develop alternative methods during the discovery phase to enhance both the efficiency of target discovery and the effectiveness of the targets identified.

The discovery phase in the drug development pipeline for combination therapy consists of two main components: (a) identifying relevant target combinations and (b) developing therapeutic compounds (\eg drugs) that act on these targets. With the rapid accumulation of experimental and omics data from HTS and the increasing availability of disease-related signaling networks, there is a growing focus on data-driven, \textit{in silico} techniques to predict effective target combinations~\cite{chua2017synergistic,HC+19,Alvarez16,TP21,WS13,li18,Chua2012,timma}. Most of these methods aim to identify a set of targets that maximizes control over deregulated genes while minimizing control over unperturbed genes in a disease context. This approach could enhance the discovery of superior drug combinations, as the identified targets can be candidates for their potential to guide the network from a disease state to a healthy one. Thus, \textit{in silico} target combination discovery (a.k.a multi-target selection) has the potential to serve as a powerful discovery and pre-screening platform when integrated with complementary technologies like HTS, ultimately reducing the time and cost of the drug discovery process~\cite{CP13,HW11,Mitchell03}.

Current state-of-the-art \textit{in silico} target combination techniques do not take into account the characteristics of approved target combinations. A significant challenge for these methods is the computational difficulty of evaluating all $k$-combinations of nodes in large signaling networks containing thousands of nodes. Additionally, there is limited research on systematically guiding the selection of the most promising node combination candidates for further analysis. Existing approaches typically tackle this intractability challenge in two ways. First, they often focus on small signaling networks with only tens to hundreds of nodes~\cite{Chua2012,chua2017synergistic,TP21,WS13,timma}, such as the 167-node cancer signaling network examined in~\cite{TP21}, which restricts the discovery of novel target combinations. Second, they simplify the input network to make the problem more manageable. For instance, Hu et al.~\cite{HC+19} convert the network into a bipartite graph and analyze 1,000 different \textit{structural control configuration} (\ie a spanning graph with the same node set) to identify \textit{optimal control nodes}. However, this simplification process is not informed by the characteristics of known drug target combinations for specific cancers. Therefore, it is crucial to develop profile-driven flexible techniques that can systematically guide the selection of candidate node combinations in large signaling networks tailored to specific cancer types for more effective target combination discovery.

\end{document}